\documentclass[english]{article}
\usepackage[T1]{fontenc}
\usepackage[latin9]{inputenc}
\usepackage{amssymb}
\usepackage{setspace}
\usepackage{esint}

\makeatletter

\let\SF@@footnote\footnote
\def\footnote{\ifx\protect\@typeset@protect
    \expandafter\SF@@footnote
  \else
    \expandafter\SF@gobble@opt
  \fi
}
\expandafter\def\csname SF@gobble@opt \endcsname{\@ifnextchar[%]
  \SF@gobble@twobracket
  \@gobble
}
\edef\SF@gobble@opt{\noexpand\protect
  \expandafter\noexpand\csname SF@gobble@opt \endcsname}
\def\SF@gobble@twobracket[#1]#2{}

\newcommand{\lyxaddress}[1]{
\par {\raggedright #1
\vspace{1.4em}
\noindent\par}
}

\makeatother

\usepackage{babel}
\begin{document}

\title{\textbf{Unifying the dynamical effects of quantum and classical noises
}\thanks{This is a first draft of the manuscript.}}

\author{Li Yu}
\maketitle

\lyxaddress{\begin{center}
\emph{Department of Physics, Harvard University, Cambridge, MA 02138,
USA}
\par\end{center}}
\begin{abstract}
We develop a new master equation as a unified description of the effects
of both quantum noise (system-bath interaction) and classical noise
on a system's dynamics, using a two-dimensional series expansion method.
When quantum and classical noises are both present, their combined
effect on a system's dynamics is not necessarily a simple sum of the
two individual effects. Thus previous master equations for open systems
and those for classical noise, even when jointly used, may not capture
the full physics. Our formalism can determine whether there is interference
between quantum and classical noises and will be able to capture and
describe such interference if there is any (in a perturbative manner).
We find that, interestingly, second-order interference between quantum
and classical noises vanishes identically. This work thus also serves
to justify simple additive treatments of quantum and classical noises,
especially in the weak coupling regime. For a Zeeman-splitted atom
in a stochastic magnetic field interacting with an optical cavity,
we use the formalism developed herein to find the overall decoherence
rate between the atom's energy levels.
\end{abstract}

\section{Introduction}

\begin{singlespace}
Ideally, for a closed quantum system under a deterministic Hamiltonian,
its density matrix evolves unitarily according to the von Newmann
equation, \cite{breuer book} 
\begin{equation}
\frac{d}{dt}\rho(t)=-i\left[H_{deterministic}(t),\rho(t)\right].
\end{equation}
In reality, many quantum systems interact with external quantum degrees
of freedom (the ``environment''), and such system-environment interactions
generally result in non-unitary dynamics at the system's level. \cite{yu,schlosshauer book}
This is the scenario of open quantum systems. On the other hand, there
are scenarios where no external quantum degree of freedom is formally
present, but a (closed) quantum system is subject to a Hamiltonian
that is non-deterministic/stochastic across different realizations
of the system's evolution (e.g. across different runs of repeated
experiment). This stochasticity is called classical noise. \cite{schlosshauer book,james2}
For a closed quantum system under a stochastic Hamiltonian, its dynamics
is unitary in one particular realization of the system's evolution
(e.g. in a single run of the experiment). However, when we repeat
the experiment for multiple times, and if we consinder the ensemble
average of the system's statistics over multiple realizations, the
system's average density matrix generally evolves non-unitarily. \cite{schlosshauer book,james2}
\footnote{\begin{singlespace}
It may be added that system-environment interaction is sometimes loosely
referred to as ``quantum noise'' in the open quantum system scenario,
because observationally it can have similar effects on the system's
statistics like classical noise does. \cite{saira}
\end{singlespace}
}

The study of how system-environment interaction and classical noise
affect a system's dynamics is important. Put most simplistically,
quantum coherence is essential to the broad field of quantum information
\cite{nielson chuang book} and quantum control \cite{brumer book}.
Two conceptually different sources for the loss of quantum coherence
are: (a) decoherence, which generally arises in open quantum systems,
resulting from system-environment interaction; \cite{yu,schlosshauer book}
(b) dephasing, which generally arises under stochastic Hamiltonians,
resulting from classical noise. \cite{james2,schlosshauer book} Observationally,
both may be similarly represented by the decay of some off-diagonal
density matrix element(s) in some bases. \cite{schlosshauer book,james2,yu}
However, the two kinds of ``noises'' are of different natures -
a system can get entangled with the environment in the case of open
quantum systems, whereas there is no system-environment entanglement
in the case of classical noise. \cite{schlosshauer book}

Open quantum systems are extensively studied in the literature. \cite{yu,breuer,breuer book,yamaguchi,elenewski,jeske,kirton}
There are also studies on quantum systems under classical noise. \cite{james2,budini}
However, there are few studies that consider the effects of both system-environment
interaction and classical noise on the system's dynamics in a unified
and systematic way. (See \cite{attal,bardet} for previous works on
classical and quantum noises, through ``Environment Algebra'' and
quantum Langevin equations rather than master equations.) In our work,
with a two-dimensional series expansion approach, we develop a master
equation formalism that takes into account both system-environment
interaction (``quantum noise'') and stochastic Hamiltonian (classical
noise) and treats their joint effects on a system's dynamics in a
unified and consistent manner.

When both quantum noise and classical noise are present, there may
be interference between the two kinds of noises on a system's dynamics.
Most cautiously put, we have no a priori reason to think that their
joint effect is merely a simple sum of the two individual effects.
If interference exists, then the master equations for open quantum
systems and those for classical noise, even when both are used together,
do not describe the full physics. The master equation formalism developed
herein will be able to determine whether there is interference between
quantum and classical noises on a system's dynamics (in a perturbative
manner). If there is interference (in some perturbative order), our
formalism can capture and quantify such interference; if there is
no interference (in some perturbative order), our formalism can rule
it out. This is a motivation behind this work.
\end{singlespace}

\section{Theory}

\subsection{Derivations \protect\footnote{All works herein are within the interaction picture unless otherwise
noted.}}

\subsubsection*{Total unitary dynamics}

The total interaction Hamiltonian consists of two terms
\begin{equation}
H_{int}^{(j)}(t)=\lambda H_{SE}(t)+\delta H_{S}^{(j)}(t)\otimes\mathbb{I}_{E},\label{eq:formal H}
\end{equation}
where $H_{SE}(t)$ is the system-bath interaction Hamiltonian, $H_{S}^{(j)}(t)\otimes\mathbb{I}_{E}$
is the stochastic Hamiltonian acting on the system, $\lambda$ and
$\delta$ parametrize the strength of the system-bath interaction
and that of the stochastic Hamiltonian respectively, and the index
$j$ denotes the $j$-th realization of the stochastic process / experimental
run. The first term alone can be lead to system-bath entanglement,
which in turn can lead to decoherence in the system's reduced density
matrix. This often goes by the name of ``quantum noise'' and is
the subject of open quantum systems. The second term alone, upon averaging,
can lead to dephasing of the system's density matrix, which is an
effect of ``classical noise''.

The system-bath total density matrix in the $j$-th run of the experiment
thus obeys the equation of motion
\begin{equation}
i\frac{d}{dt}\rho_{total}^{(j)}(t)=\left[H_{int}^{(j)}(t),\;\rho_{total}^{(j)}(t)\right]=\left[\lambda H_{SE}(t)+\delta H_{S}^{(j)}(t)\otimes\mathbb{I}_{E},\;\rho_{total}^{(j)}(t)\right].
\end{equation}

Let $U^{(j)}(t,0)$ be the unitary evolution operator for the system-bath
total density matrix in the in the $j$-th run of the experiment such
that $\rho_{total}^{(j)}(t)=U^{(j)}(t,0)\rho_{total}(0)U^{(j)\dagger}(t,0)$,
then 
\begin{equation}
i\frac{d}{dt}U^{(j)}(t,0)=\left(\lambda H_{SE}(t)+\delta H_{S}^{(j)}(t)\otimes\mathbb{I}_{E}\right)U^{(j)}(t,0).\label{eq:total u eom}
\end{equation}

\subsubsection*{Two-dimensional series expansion}

The first key step in our contruction is to suppose that the total
unitary operator $U^{(j)}(t,0)$ can be expanded in a 2-dimensional
power series of $\lambda$ and $\delta$: 
\begin{equation}
U^{(j)}(t,0)=\sum_{m=0}^{\infty}\sum_{n=0}^{\infty}\lambda^{m}\delta^{n}U_{m,n}^{(j)}(t,0).\label{eq:u 2d series}
\end{equation}

Plugging Eq.(\ref{eq:u 2d series}) into Eq.(\ref{eq:total u eom}),
we have 
\begin{eqnarray}
 &  & i\sum_{m=0}^{\infty}\sum_{n=0}^{\infty}\lambda^{m}\delta^{n}\frac{d}{dt}U_{m,n}^{(j)}(t,0)\nonumber \\
 & = & \sum_{m=0}^{\infty}\sum_{n=0}^{\infty}\left(\lambda^{m+1}\delta^{n}H_{SE}(t)U_{m,n}^{(j)}(t,0)+\lambda^{m}\delta^{n+1}H_{S}^{(j)}(t)\otimes\mathbb{I}_{E}U_{m,n}^{(j)}(t,0)\right).\label{eq:2d series eom}
\end{eqnarray}

Comparing like-order terms from the left-hand side and right-hand
side of Eq.(\ref{eq:2d series eom}), we have iterative equations
for $U_{M,N}^{(j)}(t,0)$:
\begin{eqnarray}
i\frac{d}{dt}U_{0,0}^{(j)}(t,0) & = & 0,\;\;\left(M=0,\;N=0\right)\\
i\frac{d}{dt}U_{M,0}^{(j)}(t,0) & = & H_{SE}(t)U_{M-1,0}^{(j)}(t,0),\;\;\left(M\geqslant1,\;N=0\right)\\
i\frac{d}{dt}U_{0,N}^{(j)}(t,0) & = & H_{S}^{(j)}(t)\otimes\mathbb{I}_{E}U_{0,N-1}^{(j)}(t,0),\;\;\left(M=0,\;N\geqslant1\right)\\
i\frac{d}{dt}U_{M,N}^{(j)}(t,0) & = & H_{SE}(t)U_{M-1,N}^{(j)}(t,0)+\left(H_{S}^{(j)}(t)\otimes\mathbb{I}_{E}\right)U_{M,N-1}^{(j)}(t,0).\;\;\left(M,\,N\geqslant1\right)\nonumber \\
\end{eqnarray}

Solving the above iterative equations, we have, for example, 
\begin{eqnarray}
U_{0,0}^{(j)}(t,0) & = & \mathbb{I},\\
U_{1,0}^{(j)}(t,0) & = & (-i)\int_{0}^{t}dt'H_{SE}(t'),\label{eq:U10}\\
U_{2,0}^{(j)}(t,0) & = & -\int_{0}^{t}dt'\int_{0}^{t'}dt"H_{SE}(t')H_{SE}(t"),\label{eq:U20}\\
U_{0,1}^{(j)}(t,0) & = & (-i)\int_{0}^{t}dt'H_{S}^{(j)}(t')\otimes\mathbb{I}_{E},\label{eq:U01}\\
U_{0,2}^{(j)}(t,0) & = & -\int_{0}^{t}dt'\int_{0}^{t'}dt"\left(H_{S}^{(j)}(t')H_{S}^{(j)}(t")\right)\otimes\mathbb{I}_{E},\label{eq:U02}\\
U_{1,1}^{(j)}(t,0) & = & -\int_{0}^{t}dt'\int_{0}^{t'}dt"\left[H_{SE}(t')\left(H_{S}^{(j)}(t")\otimes\mathbb{I}_{E}\right)+\left(H_{S}^{(j)}(t')\otimes\mathbb{I}_{E}\right)H_{SE}(t")\right],\nonumber \\
 & ....\label{eq:U11}
\end{eqnarray}

\subsubsection*{Averaged reduced density matrix}

Tracing out the environmental degrees of freedom and averaging over
the stochastic process, we have the averaged reduced density matrix
that describes the measurement statistics 
\begin{equation}
\bar{\rho}_{S}(t)\equiv\lim_{R\rightarrow\infty}\frac{1}{R}\sum_{j=1}^{R}Tr_{E}\left(\rho_{total}^{(j)}(t)\right).
\end{equation}

Thus for initial seperability $\rho_{total}(0)=\rho_{S}(0)\otimes\rho_{E0}$,
we have
\begin{eqnarray}
\bar{\rho}_{S}(t) & = & \lim_{R\rightarrow\infty}\frac{1}{R}\sum_{j=1}^{R}Tr_{E}\left(U^{(j)}(t,0)\rho_{S}(0)\otimes\rho_{E0}U^{(j)\dagger}(t,0)\right)\nonumber \\
 & = & \lim_{R\rightarrow\infty}\frac{1}{R}\sum_{j=1}^{R}Tr_{E}\left(\sum_{m=0}^{\infty}\sum_{n=0}^{\infty}\lambda^{m}\delta^{n}U_{m,n}^{(j)}(t,0)\rho_{S}(0)\otimes\rho_{E0}\sum_{m'=0}^{\infty}\sum_{n'=0}^{\infty}\lambda^{m'}\delta^{n'}U_{m',n'}^{(j)\dagger}(t,0)\right)\nonumber \\
 & = & \sum_{m=0}^{\infty}\sum_{n=0}^{\infty}\sum_{m'=0}^{\infty}\sum_{n'=0}^{\infty}\lambda^{m+m'}\delta^{n+n'}\lim_{R\rightarrow\infty}\frac{1}{R}\sum_{j=1}^{R}Tr_{E}\left(U_{m,n}^{(j)}(t,0)\rho_{S}(0)\otimes\rho_{E0}U_{m',n'}^{(j)\dagger}(t,0)\right)\nonumber \\
 & = & \sum_{M=0}^{\infty}\sum_{N=0}^{\infty}\lambda^{M}\delta^{N}\sum_{m=0}^{M}\sum_{n=0}^{N}\lim_{R\rightarrow\infty}\frac{1}{R}\sum_{j=1}^{R}Tr_{E}\left(U_{M-m,N-n}^{(j)}(t,0)\rho_{S}(0)\otimes\rho_{E0}U_{m,n}^{(j)\dagger}(t,0)\right).\nonumber \\
\end{eqnarray}

With $\rho_{t}\equiv\bar{\rho}_{S}(t)$ and $\rho_{0}\equiv\rho_{S}(0)$,
the linear mapping from $\rho_{0}$ to $\rho_{t}$ can be re-written
as 
\begin{eqnarray}
\rho_{t} & = & \left(\mathbb{I}+\mathcal{E}_{t}\right)(\rho_{0})\nonumber \\
 & = & \mathbb{I}(\rho_{0})+\sum_{(M,N)\neq(0,0)}\lambda^{M}\delta^{N}\mathcal{E}_{t\,(M,N)}(\rho_{0}),
\end{eqnarray}
where the $(M,N)$-th order linear map for an arbitrary system density
matrix $\rho$ is 
\begin{equation}
\mathcal{E}_{t\,(M,N)}(\rho)\equiv\sum_{m=0}^{M}\sum_{n=0}^{N}\lim_{R\rightarrow\infty}\frac{1}{R}\sum_{j=1}^{R}Tr_{E}\left(U_{M-m,N-n}^{(j)}(t,0)\rho\otimes\rho_{E0}U_{m,n}^{(j)\dagger}(t,0)\right),\label{eq:eps m,n}
\end{equation}
and 
\begin{equation}
\mathcal{E}_{t}(\rho)\equiv\sum_{(M,N)\neq(0,0)}\lambda^{M}\delta^{N}\mathcal{E}_{t\,(M,N)}(\rho).\label{eq:eps total}
\end{equation}

\subsubsection*{The $Y_{Q,t}$ map}

The second key step is to introduce a linear map that will be central
to our construction 
\begin{equation}
Y_{Q,t}(\rho)\equiv\sum_{q=0}^{Q}(-1)^{q}\mathcal{E}_{t}^{(q)}(\rho),
\end{equation}
where $\mathfrak{\mathcal{E}}_{t}^{(q)}\left(\rho\right)\equiv\mathfrak{\mathcal{E}}_{t}\left(\mathfrak{\mathcal{E}}_{t}\left(...\mathfrak{\mathcal{E}}_{t}\left(\rho\right)\right)\right)$
is a composition of $q$ $\mathfrak{\mathcal{E}}_{t}$ maps.

Applying this linear map to the averaged reduced density matrix at
time t yields 
\begin{eqnarray}
Y_{Q,t}\left(\rho_{t}\right) & = & \sum_{q=0}^{Q}(-1)^{q}\mathfrak{\mathcal{E}}_{t}^{(q)}\left(\left(\mathbb{I}+\mathfrak{\mathcal{E}}_{t}\right)\left(\rho_{0}\right)\right)\nonumber \\
 & = & \mathbb{I}\left(\left(\mathbb{I}+\mathfrak{\mathcal{E}}_{t}\right)\left(\rho_{0}\right)\right)-\mathfrak{\mathcal{E}}_{t}\left(\left(\mathbb{I}+\mathfrak{\mathcal{E}}_{t}\right)\left(\rho_{0}\right)\right)+\mathfrak{\mathcal{E}}_{t}\left(\mathfrak{\mathcal{E}}_{t}\left(\left(\mathbb{I}+\mathfrak{\mathcal{E}}_{t}\right)\left(\rho_{0}\right)\right)\right)-....\nonumber \\
 & = & \mathbb{I}\left(\rho_{0}\right)+\mathfrak{\mathcal{E}}_{t}\left(\rho_{0}\right)-\mathfrak{\mathcal{E}}_{t}\left(\rho_{0}\right)-\mathfrak{\mathcal{E}}_{t}\left(\mathfrak{\mathcal{E}}_{t}\left(\rho_{0}\right)\right)+\mathfrak{\mathcal{E}}_{t}\left(\mathfrak{\mathcal{E}}_{t}\left(\rho_{0}\right)\right)+\mathfrak{\mathcal{E}}_{t}\left(\mathfrak{\mathcal{E}}_{t}\left(\mathfrak{\mathcal{E}}_{t}\left(\rho_{0}\right)\right)\right)-....\nonumber \\
 & = & \left(\mathbb{I}+(-1)^{Q}\mathfrak{\mathcal{E}}_{t}^{(Q+1)}\right)\left(\rho_{0}\right).
\end{eqnarray}
Rearranging the terms, we now have the crucial equality in our work:
\begin{eqnarray}
\rho_{0} & = & Y_{Q,t}\left(\rho_{t}\right)+(-1)^{Q+1}\mathfrak{\mathcal{E}}_{t}^{(Q+1)}\left(\rho_{0}\right)\nonumber \\
 & = & \sum_{q=0}^{Q}(-1)^{q}\mathfrak{\mathcal{E}}_{t}^{(q)}\left(\rho_{t}\right)+(-1)^{Q+1}\mathfrak{\mathcal{E}}_{t}^{(Q+1)}\left(\rho_{0}\right).\label{eq:inverse relation}
\end{eqnarray}
What it does is to express the initial state $\rho_{0}$ in terms
of the state at time t $\rho_{t}$ (with residual term on $\rho_{0}$
that can be negleted to certain perturbative order).

Before proceeding further, let's examine the order of magnitudes of
relevant terms. For the 1st-order term:
\begin{eqnarray}
\mathcal{E}_{t}(\rho) & = & \lambda\mathcal{E}_{t\,(1,0)}(\rho)+\delta\mathcal{E}_{t\,(0,1)}(\rho)+\lambda\delta\mathcal{E}_{t\,(1,1)}(\rho)+....\nonumber \\
 & = & \mathcal{O}(\lambda)+\mathcal{O}(\delta);
\end{eqnarray}
For the 2nd-order term:
\begin{eqnarray}
\mathcal{E}_{t}^{(2)}(\rho)=\mathcal{E}_{t}\left(\mathcal{E}_{t}(\rho)\right) & = & \mathcal{E}_{t}\left(\lambda\mathcal{E}_{t\,(1,0)}(\rho)+\delta\mathcal{E}_{t\,(0,1)}(\rho)+....\right)\nonumber \\
 & = & \lambda\mathcal{E}_{t\,(1,0)}\left(\lambda\mathcal{E}_{t\,(1,0)}(\rho)+\delta\mathcal{E}_{t\,(0,1)}(\rho)+....\right)\nonumber \\
 &  & +\delta\mathcal{E}_{t\,(0,1)}\left(\lambda\mathcal{E}_{t\,(1,0)}(\rho)+\delta\mathcal{E}_{t\,(0,1)}(\rho)+....\right)+....\nonumber \\
 & = & \lambda^{2}\mathcal{E}_{t\,(1,0)}\left(\mathcal{E}_{t\,(1,0)}(\rho)\right)+\lambda\delta\mathcal{E}_{t\,(1,0)}\left(\mathcal{E}_{t\,(0,1)}(\rho)\right)\nonumber \\
 &  & +\lambda\delta\mathcal{E}_{t\,(0,1)}\left(\mathcal{E}_{t\,(1,0)}(\rho)\right)+\delta^{2}\mathcal{E}_{t\,(0,1)}\left(\delta\mathcal{E}_{t\,(0,1)}(\rho)\right)+....\nonumber \\
 & = & \mathcal{O}(\lambda^{2})+\mathcal{O}(\lambda\delta)+\mathcal{O}(\delta^{2});
\end{eqnarray}
In general, for the $Q$th-order term:
\begin{eqnarray}
\mathcal{E}_{t}^{(Q)}(\rho) & = & \sum_{q=0}^{Q}\mathcal{O}\left(\lambda^{q}\delta^{Q-q}\right).\label{eq:order of magnitude}
\end{eqnarray}

\subsubsection*{Equation of motion }

\begin{doublespace}
Differentiating the averaged reduced density matrix with respect to
time, we have
\begin{eqnarray}
\frac{d}{dt}\rho_{t} & = & \frac{d}{dt}\left(\mathbb{I}+\mathfrak{\mathcal{E}}_{t}\right)\left(\rho_{0}\right)\nonumber \\
 & = & \dot{\mathfrak{\mathcal{E}_{t}}}\left(\rho_{0}\right)\nonumber \\
 & = & \dot{\mathfrak{\mathcal{E}_{t}}}\left(\sum_{q=0}^{Q}(-1)^{q}\mathfrak{\mathcal{E}}_{t}^{(q)}\left(\rho_{t}\right)+(-1)^{Q+1}\mathfrak{\mathcal{E}}_{t}^{(Q+1)}\left(\rho_{0}\right)\right),
\end{eqnarray}
where in the last equality we have made use of Eq.(\ref{eq:inverse relation}).
Thus we now have 
\begin{equation}
\frac{d}{dt}\rho_{t}=\sum_{q=0}^{Q}(-1)^{q}\dot{\mathfrak{\mathcal{E}_{t}}}\left(\mathfrak{\mathcal{E}}_{t}^{(q)}\left(\rho_{t}\right)\right)+(-1)^{Q+1}\dot{\mathfrak{\mathcal{E}_{t}}}\left(\mathfrak{\mathcal{E}}_{t}^{(Q+1)}\left(\rho_{0}\right)\right).\label{eq:eom inhomo}
\end{equation}
Note that no approximation has been made so far and that Eq.(\ref{eq:eom inhomo})
is formally exact.
\end{doublespace}

\begin{singlespace}
With Eq.(\ref{eq:eom inhomo}), we can systematically make approximations,
that is, collecting like-order terms in $\lambda$ and $\delta$ and
truncate the series as needed. For example, suppose we want to consider
$P$th-order approximation (i.e. approximations up to $\sum_{q=0}^{P}\mathcal{O}\left(\lambda^{q}\delta^{P-q}\right)$
terms). Because as we have shown in Eq.(\ref{eq:order of magnitude})
$\dot{\mathfrak{\mathcal{E}}_{t}}\left(\mathfrak{\mathcal{E}}_{t}^{(Q+1)}\left(\rho_{0}\right)\right)\sim\sum_{q=0}^{Q+2}\mathcal{O}\left(\lambda^{q}\delta^{Q+2-q}\right)$,
we can always choose $Q\geqslant P-1$, so that the residual term
$(-1)^{Q+1}\dot{\mathfrak{\mathcal{E}_{t}}}\left(\mathfrak{\mathcal{E}}_{t}^{(Q+1)}\left(\rho_{0}\right)\right)$
is negligible to our intended approximation. \footnote{Note that ``$P$th-order approximation'' in this context has a slightly
different meaning than ``$P$th-order approximation'' in the case
of a 1-dimensional series expansion (e.g. in \cite{yu}). In the case
of a 2-dimensional series expansion, a ``$P$th-order approximation''
includes all terms with total power of $P$, namely $\lambda^{P}$,
$\lambda^{P-1}\delta$, $\lambda^{P-2}\delta^{2}$, and so on.} \footnote{As a side note, our formalism would also allow for truncation of the
2-dimensional series to, say $\mathcal{O}\left(\lambda^{M}\delta^{N}\right)$,
for arbitrary $\left(M,\,N\right)$ of our choice. For example, if
the system-bath interaction effect is more significant and the classical
noise effect is comparatively less important, we may truncate to the
2-dimensional series to a higher order in $\lambda^{M}$ and lower
order in $\delta^{N}$, that is, for $M>N$.} 

A formally exact, time-local equation of motion can be formally achieved
by taking the $Q\rightarrow\infty$ limit on the right-hand side of
Eq.(\ref{eq:eom inhomo}). Loosely speaking, as $\lim_{Q\rightarrow\infty}(-1)^{Q+1}\dot{\mathfrak{\mathcal{E}}_{t}}\left(\mathfrak{\mathcal{E}}_{t}^{(Q+1)}\left(\rho_{0}\right)\right)\sim\lim_{Q\rightarrow\infty}\sum_{q=0}^{Q+2}\mathcal{O}\left(\lambda^{q}\delta^{Q+2-q}\right)\rightarrow0$,
the residual term can be neglected, and we obtain 
\begin{equation}
\frac{d}{dt}\rho_{t}=\sum_{q=0}^{\infty}(-1)^{q}\dot{\mathfrak{\mathcal{E}_{t}}}\left(\mathfrak{\mathcal{E}}_{t}^{(q)}\left(\rho_{t}\right)\right).
\end{equation}
This $Q\rightarrow\infty$ formal treatment and the resulting equation
of motion are not necessary in practice, however. See \cite{yu} for
discussions on this issue.
\end{singlespace}

\subsection{Equation of motion}

\subsubsection{Second-order equation of motion}

To work out the second-order equation of motion for the average reduced
density matrix, we first set $Q=1$ in Eq.(\ref{eq:eom inhomo}):
\begin{equation}
\frac{d}{dt}\rho_{t}=\dot{\mathfrak{\mathcal{E}_{t}}}\left(\rho_{t}\right)-\dot{\mathfrak{\mathcal{E}_{t}}}\left(\mathfrak{\mathcal{E}}_{t}\left(\rho_{t}\right)\right)+\dot{\mathfrak{\mathcal{E}_{t}}}\left(\mathfrak{\mathcal{E}}_{t}^{(2)}\left(\rho_{0}\right)\right),
\end{equation}
so that the residual term $\dot{\mathfrak{\mathcal{E}_{t}}}\left(\mathfrak{\mathcal{E}}_{t}^{(2)}\left(\rho_{0}\right)\right)\sim\sum_{q=0}^{3}\mathcal{O}\left(\lambda^{q}\delta^{3-q}\right)$
is of third order significance and thus negligible in second-order
approximation. Thus up to second order we have 
\begin{eqnarray}
\frac{d}{dt}\rho_{t} & = & \dot{\mathfrak{\mathcal{E}_{t}}}\left(\rho_{t}\right)-\dot{\mathfrak{\mathcal{E}_{t}}}\left(\mathfrak{\mathcal{E}}_{t}\left(\rho_{t}\right)\right)\nonumber \\
 & = & \lambda\dot{\mathcal{E}}_{t\,(1,0)}(\rho_{t})+\delta\dot{\mathcal{E}}_{t\,(0,1)}(\rho_{t})+\lambda^{2}\dot{\mathcal{E}}_{t\,(2,0)}(\rho_{t})+\lambda\delta\dot{\mathcal{E}}_{t\,(1,1)}(\rho_{t})+\delta^{2}\dot{\mathcal{E}}_{t\,(0,2)}(\rho_{t})+....\nonumber \\
 &  & -\lambda\dot{\mathcal{E}}_{t\,(1,0)}\left(\lambda\mathcal{E}_{t\,(1,0)}(\rho_{t})+\delta\mathcal{E}_{t\,(0,1)}(\rho_{t})+....\right)\nonumber \\
 &  & -\delta\dot{\mathcal{E}}_{t\,(0,1)}\left(\lambda\mathcal{E}_{t\,(1,0)}(\rho_{t})+\delta\mathcal{E}_{t\,(0,1)}(\rho_{t})+....\right)+....
\end{eqnarray}
where we have neglected all terms of third or higher orders.

Collecting like-order terms, we obtain the equation of motion for
the ``average reduced density matrix'' $\rho_{t}$ of the system
(up to second order) 
\begin{equation}
\frac{d}{dt}\rho_{t}=\lambda\mathcal{L}_{t\,(1,0)}(\rho_{t})+\delta\mathcal{L}_{t\,(0,1)}(\rho_{t})+\lambda^{2}\mathcal{L}_{t\,(2,0)}(\rho_{t})+\delta^{2}\mathcal{L}_{t\,(0,2)}(\rho_{t})+\lambda\delta\mathcal{L}_{t\,(1,1)}(\rho_{t}),\label{eq:2nd order eom general}
\end{equation}
where
\begin{eqnarray}
\mathcal{L}_{t\,(1,0)}(\rho_{t}) & = & \dot{\mathcal{E}}_{t\,(1,0)}(\rho_{t}),\label{eq:L10}\\
\mathcal{L}_{t\,(0,1)}(\rho_{t}) & = & \dot{\mathcal{E}}_{t\,(0,1)}(\rho_{t}),\label{eq:L01}\\
\mathcal{L}_{t\,(2,0)}(\rho_{t}) & = & \dot{\mathcal{E}}_{t\,(2,0)}(\rho_{t})-\dot{\mathcal{E}}_{t\,(1,0)}\left(\mathcal{E}_{t\,(1,0)}(\rho_{t})\right),\label{eq:L20}\\
\mathcal{L}_{t\,(0,2)}(\rho_{t}) & = & \dot{\mathcal{E}}_{t\,(0,2)}(\rho_{t})-\dot{\mathcal{E}}_{t\,(0,1)}\left(\mathcal{E}_{t\,(0,1)}(\rho_{t})\right),\label{eq:L02}\\
\mathcal{L}_{t\,(1,1)}(\rho_{t}) & = & \dot{\mathcal{E}}_{t\,(1,1)}(\rho_{t})-\dot{\mathcal{E}}_{t\,(1,0)}\left(\mathcal{E}_{t\,(0,1)}(\rho_{t})\right)-\dot{\mathcal{E}}_{t\,(0,1)}\left(\mathcal{E}_{t\,(1,0)}(\rho_{t})\right),\label{eq:L11}
\end{eqnarray}
with $\mathcal{E}_{t\,(M,N)}(\rho_{t})$ defined in Eq.(\ref{eq:eps m,n}).

Even with this abstract form of the second-order equation of motion,
we can already see that $\mathcal{L}_{t\,(2,0)}(\rho_{t})$ is the
familiar decoherence term due to system-bath interaction (i.e. ``quantum
noise''), $\mathcal{L}_{t\,(0,2)}(\rho_{t})$ is the familiar dephasing
term due to classical noise, and $\mathcal{L}_{t\,(1,1)}(\rho_{t})$
supposedly represents the interference between quantum noise and classical
noise on the system's dynamics, if it does not vanish.

\subsubsection{Second-order cross term }

The lowest-order contribution of quantum noise to decoherence is $\mathcal{L}_{t\,(2,0)}(\rho_{t})$,
and the lowest-order contribution of classical noise to dephasing
is $\mathcal{L}_{t\,(0,2)}(\rho_{t})$, both of which are of second
order in nature. To this same order, a cross-term $\mathcal{L}_{t\,(1,1)}(\rho_{t})$
as defined by Eq.(\ref{eq:L11}) originates neither from quantum noise
alone nor from classical noise alone, but supposedly from both.

To work out the details of $\mathcal{L}_{t\,(1,1)}(\rho_{t})$ as
in Eq.(\ref{eq:L11}), we first make use of Eq.(\ref{eq:eps m,n})
for relevant values of $(m,n)$ to work out various individual terms:
\begin{eqnarray}
\mathcal{E}_{t\,(1,0)}(\rho) & = & \lim_{R\rightarrow\infty}\frac{1}{R}\sum_{j=1}^{R}Tr_{E}\left(U_{1,0}^{(j)}(t,0)\rho\otimes\rho_{E0}+\rho\otimes\rho_{E0}U_{1,0}^{(j)\dagger}(t,0)\right)\nonumber \\
 & = & (-i)\int_{0}^{t}dt'Tr_{E}\left(H_{SE}(t')\rho\otimes\rho_{E0}-\rho\otimes\rho_{E0}H_{SE}(t')\right)\nonumber \\
 & = & (-i)\int_{0}^{t}dt'Tr_{E}\left(\left[H_{SE}(t'),\;\rho\otimes\rho_{E0}\right]\right),\label{eq:eps 1,0}\\
\Rightarrow\:\dot{\mathcal{E}}_{t\,(1,0)}(\rho) & = & (-i)Tr_{E}\left(\left[H_{SE}(t),\;\rho\otimes\rho_{E0}\right]\right);\label{eq:eps 1,0 dot}
\end{eqnarray}
\begin{eqnarray}
\mathcal{E}_{t\,(0,1)}(\rho) & = & \lim_{R\rightarrow\infty}\frac{1}{R}\sum_{j=1}^{R}Tr_{E}\left(U_{0,1}^{(j)}(t,0)\rho\otimes\rho_{E0}+\rho\otimes\rho_{E0}U_{0,1}^{(j)\dagger}(t,0)\right)\nonumber \\
 & = & (-i)\int_{0}^{t}dt'\lim_{R\rightarrow\infty}\frac{1}{R}\sum_{j=1}^{R}Tr_{E}\left(\left(H_{S}^{(j)}(t')\rho\right)\otimes\rho_{E0}-\left(\rho H_{S}^{(j)}(t')\right)\otimes\rho_{E0}\right)\nonumber \\
 & = & (-i)\int_{0}^{t}dt'\lim_{R\rightarrow\infty}\frac{1}{R}\sum_{j=1}^{R}\left[H_{S}^{(j)}(t'),\;\rho\right]\nonumber \\
 & = & (-i)\int_{0}^{t}dt'\left[\overline{H_{S}(t')},\;\rho\right],\label{eq:eps 0,1}\\
\Rightarrow\:\dot{\mathcal{E}}_{t\,(0,1)}(\rho) & = & (-i)\left[\overline{H_{S}(t)},\;\rho\right],\label{eq:eps 0,1 dot}
\end{eqnarray}
where the statistical average is defined as $\overline{H_{S}(t)}\equiv\lim_{R\rightarrow\infty}\frac{1}{R}\sum_{j=1}^{R}H_{S}^{(j)}(t)$;
\begin{eqnarray}
\mathcal{E}_{t\,(1,1)}(\rho) & = & \lim_{R\rightarrow\infty}\frac{1}{R}\sum_{j=1}^{R}Tr_{E}\{\,U_{1,1}^{(j)}(t,0)\rho\otimes\rho_{E0}+\rho\otimes\rho_{E0}U_{1,1}^{(j)\dagger}(t,0)\nonumber \\
 &  & +U_{1,0}^{(j)}(t,0)\rho\otimes\rho_{E0}U_{0,1}^{(j)\dagger}(t,0)+U_{0,1}^{(j)}(t,0)\rho\otimes\rho_{E0}U_{1,0}^{(j)\dagger}(t,0)\,\},\label{eq:eps 1,1}\\
\Rightarrow\:\dot{\mathcal{E}}_{t\,(1,1)}(\rho) & = & \lim_{R\rightarrow\infty}\frac{1}{R}\sum_{j=1}^{R}Tr_{E}\{\,\dot{U}_{1,1}^{(j)}(t,0)\rho\otimes\rho_{E0}+\rho\otimes\rho_{E0}\dot{U}_{1,1}^{(j)\dagger}(t,0)\nonumber \\
 &  & +\dot{U}_{1,0}^{(j)}(t,0)\rho\otimes\rho_{E0}U_{0,1}^{(j)\dagger}(t,0)+U_{1,0}^{(j)}(t,0)\rho\otimes\rho_{E0}\dot{U}_{0,1}^{(j)\dagger}(t,0)\nonumber \\
 &  & +\dot{U}_{0,1}^{(j)}(t,0)\rho\otimes\rho_{E0}U_{1,0}^{(j)\dagger}(t,0)+U_{0,1}^{(j)}(t,0)\rho\otimes\rho_{E0}\dot{U}_{1,0}^{(j)\dagger}(t,0)\,\}\nonumber \\
 & = & \int_{0}^{t}dt'\:Tr_{E}\{\,-H_{SE}(t)\left(\overline{H_{S}(t')}\otimes\mathbb{I}_{E}\right)\left(\rho\otimes\rho_{E0}\right)-\left(\overline{H_{S}(t)}\otimes\mathbb{I}_{E}\right)H_{SE}(t')\left(\rho\otimes\rho_{E0}\right)\nonumber \\
 &  & -\left(\rho\otimes\rho_{E0}\right)\left(\overline{H_{S}(t')}\otimes\mathbb{I}_{E}\right)H_{SE}(t)-\left(\rho\otimes\rho_{E0}\right)H_{SE}(t')\left(\overline{H_{S}(t)}\otimes\mathbb{I}_{E}\right)\nonumber \\
 &  & +H_{SE}(t)\left(\rho\otimes\rho_{E0}\right)\left(\overline{H_{S}(t')}\otimes\mathbb{I}_{E}\right)+H_{SE}(t')\left(\rho\otimes\rho_{E0}\right)\left(\overline{H_{S}(t)}\otimes\mathbb{I}_{E}\right)\nonumber \\
 &  & +\left(\overline{H_{S}(t)}\otimes\mathbb{I}_{E}\right)\left(\rho\otimes\rho_{E0}\right)H_{SE}(t')+\left(\overline{H_{S}(t')}\otimes\mathbb{I}_{E}\right)\left(\rho\otimes\rho_{E0}\right)H_{SE}(t)\:\}.\label{eq:eps 1,1 dot}
\end{eqnarray}

With these results, one will be able to evaluate the second-order
cross term
\begin{equation}
\mathcal{L}_{t\,(1,1)}(\rho_{t})=\dot{\mathcal{E}}_{t\,(1,1)}(\rho_{t})-\dot{\mathcal{E}}_{t\,(1,0)}\left(\mathcal{E}_{t\,(0,1)}(\rho_{t})\right)-\dot{\mathcal{E}}_{t\,(0,1)}\left(\mathcal{E}_{t\,(1,0)}(\rho_{t})\right),
\end{equation}
which will show if and how quantum and classical noises interfere
on the system's dynamics up to second order.

\subsubsection{Second-order non-interference}

Plugging Eqs.(\ref{eq:eps 1,0}-\ref{eq:eps 1,1 dot}) into Eq.(\ref{eq:L11}),
we can now calculate the cross term $\mathcal{L}_{t\,(1,1)}(\rho_{t})$.
We will work out the details term by term. The first term $\dot{\mathcal{E}}_{t\,(1,1)}(\rho)$
is readily worked out in Eq.(\ref{eq:eps 1,1 dot}); the second term
is 
\begin{eqnarray}
 &  & -\dot{\mathcal{E}}_{t\,(1,0)}\left(\mathcal{E}_{t\,(0,1)}(\rho)\right)\nonumber \\
 & = & -(-i)Tr_{E}\left(\left[H_{SE}(t),\;\mathcal{E}_{t\,(0,1)}(\rho)\otimes\rho_{E0}\right]\right)\nonumber \\
 & = & i\,Tr_{E}\left(\left[H_{SE}(t),\;\left((-i)\int_{0}^{t}dt'\left[\overline{H_{S}(t')},\;\rho\right]\right)\otimes\rho_{E0}\right]\right)\nonumber \\
 & = & \int_{0}^{t}dt'Tr_{E}\left(\left[H_{SE}(t),\;\left(\overline{H_{S}(t')}\rho\otimes\rho_{E0}-\rho\overline{H_{S}(t')}\otimes\rho_{E0}\right)\right]\right)\nonumber \\
 & = & \int_{0}^{t}dt'Tr_{E}\{\,H_{SE}(t)\left(\overline{H_{S}(t')}\otimes\mathbb{I}_{E}\right)\left(\rho\otimes\rho_{E0}\right)-H_{SE}(t)\left(\rho\otimes\rho_{E0}\right)\left(\overline{H_{S}(t')}\otimes\mathbb{I}_{E}\right)\nonumber \\
 &  & -\left(\overline{H_{S}(t')}\otimes\mathbb{I}_{E}\right)\left(\rho\otimes\rho_{E0}\right)H_{SE}(t)+\left(\rho\otimes\rho_{E0}\right)\left(\overline{H_{S}(t')}\otimes\mathbb{I}_{E}\right)H_{SE}(t)\,\};\label{eq:2nd term of L11}
\end{eqnarray}
and the third term is 
\begin{eqnarray}
 &  & -\dot{\mathcal{E}}_{t\,(0,1)}\left(\mathcal{E}_{t\,(1,0)}(\rho)\right)\nonumber \\
 & = & -(-i)\left[\overline{H_{S}(t)},\:\mathcal{E}_{t\,(1,0)}(\rho)\right]\nonumber \\
 & = & i\,\left[\overline{H_{S}(t)},\:\left((-i)\int_{0}^{t}dt'Tr_{E}\left(\left[H_{SE}(t'),\;\rho\otimes\rho_{E0}\right]\right)\right)\right]\nonumber \\
 & = & \int_{0}^{t}dt'\left[\overline{H_{S}(t)},\:\left(Tr_{E}\left(H_{SE}(t')\left(\rho\otimes\rho_{E0}\right)-\left(\rho\otimes\rho_{E0}\right)H_{SE}(t')\right)\right)\right]\nonumber \\
 & = & \int_{0}^{t}dt'\{\,\overline{H_{S}(t)}Tr_{E}\left(H_{SE}(t')\left(\rho\otimes\rho_{E0}\right)\right)-\overline{H_{S}(t)}Tr_{E}\left(\left(\rho\otimes\rho_{E0}\right)H_{SE}(t')\right)\nonumber \\
 &  & -Tr_{E}\left(H_{SE}(t')\left(\rho\otimes\rho_{E0}\right)\right)\overline{H_{S}(t)}+Tr_{E}\left(\left(\rho\otimes\rho_{E0}\right)H_{SE}(t')\right)\overline{H_{S}(t)}\,\}\nonumber \\
 & = & \int_{0}^{t}dt'Tr_{E}\{\,\left(\overline{H_{S}(t)}\otimes\mathbb{I}_{E}\right)H_{SE}(t')\left(\rho\otimes\rho_{E0}\right)-\left(\overline{H_{S}(t)}\otimes\mathbb{I}_{E}\right)\left(\rho\otimes\rho_{E0}\right)H_{SE}(t')\nonumber \\
 &  & -H_{SE}(t')\left(\rho\otimes\rho_{E0}\right)\left(\overline{H_{S}(t)}\otimes\mathbb{I}_{E}\right)+\left(\rho\otimes\rho_{E0}\right)H_{SE}(t')\left(\overline{H_{S}(t)}\otimes\mathbb{I}_{E}\right)\,\}.\label{eq:3rd term of L11}
\end{eqnarray}
Comparing the three terms Eqs.(\ref{eq:eps 1,1 dot}, \ref{eq:2nd term of L11},
\ref{eq:3rd term of L11}) of $\mathcal{L}_{t\,(1,1)}(\rho_{t})$,
we can see that they cancel out, that is, 
\begin{eqnarray}
\mathcal{L}_{t\,(1,1)}(\rho_{t}) & = & \dot{\mathcal{E}}_{t\,(1,1)}(\rho_{t})-\dot{\mathcal{E}}_{t\,(1,0)}\left(\mathcal{E}_{t\,(0,1)}(\rho_{t})\right)-\dot{\mathcal{E}}_{t\,(0,1)}\left(\mathcal{E}_{t\,(1,0)}(\rho_{t})\right)\nonumber \\
 & = & 0.\label{eq:L11=00003D=00003D0}
\end{eqnarray}

The second-order cross term is thus shown to vanish identically, that
is, free of conditions/assumptions. Physically, this means the effects
of quantum noise and classical noise on the system's average reduced
dynamics do not interfere with each other up to second order.

Because both decoherence due to quantum noise and dephasing due to
classical noise are primarily second-order effects, we may now say
that quantum and classical noises do not interfere at their dominant
order. This also implies that interference between quantum and classical
noises, if there is any, should be perturbatively less significant
than pure decoherence and pure dephasing effects. Thus this result
may also be viewed as providing justification for the practice of
treating the effects of quantum noise and classical noise in a simple
additive manner (in the weak coupling limit where second-order effects
dominate).

On the other hand, we don't have a priori reasons to think that interference
between quantum and classical noises must vanish in second order.
At least it is not obvious from the definition of the second-order
cross term $\mathcal{L}_{t\,(1,1)}(\rho_{t})$ Eq.(\ref{eq:L11})
that it should vanish identically. This second-order non-interference
is an interesting finding, the reason of which may be worth further
investigation.

\subsubsection{Higher-order dynamics}

\begin{singlespace}
Besides second-order results, the master equation formalism can be
used to systematically study a general system's average reduced dynamics
in higher orders. One can use Eqs.(\ref{eq:eom inhomo}, \ref{eq:eps total},
\ref{eq:eps m,n}) to work out higher-order equations of motion mechanically.
\end{singlespace}

In particular, our formalism can be used to work out the details of
higher-order cross terms and determine if and how quantum and classical
noises interfere in higher orders. For example, it can be shown that
an interference term in third order $\mathcal{L}_{t\,(2,1)}(\rho_{t})$
is not identically vanishing, the calculation details of which can
be found in Appendix A. This also provides evidence for interference
between quantum and classical noises on a system's dynamics.

\section{Example: a Zeeman-splitted atom in stochastic B-field interacting
with optical cavity}

In the presence of an external magnetic field, an atom can experience
the Zeeman effect, where the spacings between the splitted energy
levels are linear on the B-field strength. \cite{shankar book} Suppose
only two energy levels of the atom are relevant for the purpose of
our discussion. Now if the external B-field is stochastic instead
of deterministic, then mathematically the two-level atom is subject
to a stochastic Hamiltonian (i.e. with classical noise). \cite{james2}
This noise/stochasticity can result from the fact that experimentalists
do not have perfect control over the external B-field. Suppose further
that the aforementioned atom is also interacting with an optical cavity.
\cite{walther} In this case the atom is subject to both classical
and quantum noises at the same time, thus the formalism developed
herein can be used.

\subsection{Problem description}

\subsubsection*{Hamiltonian}

The total Hamiltonian in Schrodinger picture is 
\begin{equation}
H_{total}=\frac{\omega_{0}}{2}\sigma_{z}+\frac{a(t)}{2}\sigma_{z}+\sum_{k}\omega_{k}b_{k}^{\dagger}b_{k}+\sum_{k}g_{k}\left(\sigma_{+}b_{k}+\sigma_{-}b_{k}^{\dagger}\right),
\end{equation}
where the first term is the deterministic and time-independent part
of the two-level atom's self-Hamiltonian, $\omega_{0}$ being the
energy spacing between the two levels, the second term is the stochastic
part of the atom's self-Hamiltonian, $a(t)$ being a stochastic process
describing the fluctuating energy spacing between the two levels,
\cite{james2} the third term is the self-Hamiltonian of the optical
cavity, $\omega_{k}$ being the frequency of each cavity mode, and
the last term is the interaction between the atom and the optical
cavity, $g_{k}$ being the interaction strength. \cite{walther}

Switching to the rotating frame generated by the deterministic self-Hamiltonian
$H_{0}=\frac{\omega_{0}}{2}\sigma_{z}+\sum_{k}\omega_{k}b_{k}^{\dagger}b_{k}$,
and treating the stochastic self-Hamiltonian and the atom-cavity interaction
as a perturbing Hamiltonian $H_{perturb}=\frac{a(t)}{2}\sigma_{z}+\sum_{k}g_{k}\left(\sigma_{+}b_{k}+\sigma_{-}b_{k}^{\dagger}\right)$,
we may obtain the interaction picture Hamiltonian \cite{shankar book}
\begin{eqnarray}
H_{int} & = & e^{itH_{0}}H_{perturb}e^{-itH_{0}}\nonumber \\
 & = & \frac{a(t)}{2}\sigma_{z}+\sum_{k}\left(g_{k}e^{-i\left(\omega_{k}-\omega_{0}\right)t}\sigma_{+}b_{k}+g_{k}e^{i\left(\omega_{k}-\omega_{0}\right)t}\sigma_{-}b_{k}^{\dagger}\right).
\end{eqnarray}
Alternatively, to put it into a language conforming to our formalism
as in Eq.(\ref{eq:formal H}),
\begin{equation}
H_{int}^{(j)}=\lambda\sum_{k}\left(g_{k}e^{-i\left(\omega_{k}-\omega_{0}\right)t}\sigma_{+}b_{k}+g_{k}e^{i\left(\omega_{k}-\omega_{0}\right)t}\sigma_{-}b_{k}^{\dagger}\right)+\delta\frac{a^{(j)}(t)}{2}\sigma_{z}\otimes\mathbb{I}_{E},
\end{equation}
where $\lambda$ parametrizes the strength of the system-bath interaction
and $\delta$ parametrizes the strength of the stochastic self-Hamiltonian,
two parameters around which the two-dimensional series can be expanded,
and the index $j$ refers to the $j$-th realization of the stochastic
process. That is, we make the following identification: 
\begin{eqnarray}
H_{SE}(t) & = & \sum_{k}\left(g_{k}e^{-i\left(\omega_{k}-\omega_{0}\right)t}\sigma_{+}b_{k}+g_{k}e^{i\left(\omega_{k}-\omega_{0}\right)t}\sigma_{-}b_{k}^{\dagger}\right),\\
H_{S}^{(j)}(t) & = & \frac{a^{(j)}(t)}{2}\sigma_{z}.\label{eq:H random}
\end{eqnarray}

\subsubsection*{Initial state of cavity}

Suppose the optical cavity is initially in the thermal state, that
is, $\rho_{E0}=\frac{1}{Z}\exp\left(-\beta H_{cavity}\right)$, where
$Z=Tr_{E}\left(\exp\left(-\beta H_{cavity}\right)\right)$ is the
partition function and $\beta=1/k_{B}T$ is the inverse temperature.
\cite{schlosshauer book} In this example, we have
\begin{eqnarray}
\rho_{E0} & = & \prod_{k}\left(\frac{1}{Z_{k}}\sum_{m_{k}=0}^{\infty}e^{-m_{k}\beta\omega_{k}}|m_{k}\rangle\langle m_{k}|\right)\nonumber \\
 & = & \frac{1}{Z}\prod_{k}\left(\sum_{m_{k}=0}^{\infty}e^{-m_{k}\beta\omega_{k}}|m_{k}\rangle\langle m_{k}|\right),
\end{eqnarray}
where $Z_{k}=\sum_{m_{k}=0}^{\infty}e^{-m_{k}\beta\omega_{k}}$ and
$Z=\prod_{k}Z_{k}$, $\omega_{k}$ is the frequency of the $k$-th
cavity mode, and $m_{k}$ is the number of photons in this cavity
mode.

\subsubsection*{Stochastic property of atomic energy spacing}

Suppose the stochastic energy spacing between the two atomic levels
$a(t)$ is a real-valued Gaussian random process \cite{james2} with
a constant zero mean, \footnote{Note that any non-zero mean can be incorporated into the deterministic
part of the self-Hamiltonian and thus into $H_{0}$.}
\begin{equation}
\overline{a(t)}=0.\label{eq:zero mean}
\end{equation}

\subsection{Equation of motion}

It can be shown that the equation of motion for this physical example
up to second order is 
\begin{eqnarray}
\frac{d}{dt}\rho_{t} & = & -i\,\left[H_{eff}(t),\,\rho_{t}\right]-D_{R}(t)\left(\sigma_{-}\sigma_{+}\rho_{t}+\rho_{t}\sigma_{-}\sigma_{+}-2\sigma_{+}\rho_{t}\sigma_{-}\right)\nonumber \\
 &  & -D'_{R}(t)\left(\sigma_{+}\sigma_{-}\rho_{t}+\rho_{t}\sigma_{+}\sigma_{-}-2\sigma_{-}\rho_{t}\sigma_{+}\right)-2D_{C}(t)\left(\rho_{t}-\sigma_{z}\rho_{t}\sigma_{z}\right),\nonumber \\
\label{eq:2nd order eom example}
\end{eqnarray}
where the effective Hamiltonian is $H_{eff}(t)\equiv D_{I}(t)\sigma_{-}\sigma_{+}-D'_{I}(t)\sigma_{+}\sigma_{-}$
and the prefactors $D_{R}(t)$, $D_{I}(t)$, $D'_{R}(t)$, $D'_{I}(t)$,
and $D_{C}(t)$ are defined as 
\begin{eqnarray}
D_{R}(t) & \equiv & \int_{0}^{t}dt'\sum_{k}|g_{k}|^{2}\bar{N_{k}}\cos\left(\omega_{k0}(t-t')\right),\label{eq:Dr def}\\
D_{I}(t) & \equiv & \int_{0}^{t}dt'\sum_{k}|g_{k}|^{2}\bar{N_{k}}\sin\left(\omega_{k0}(t-t')\right),\label{eq:Di def}\\
D'_{R}(t) & \equiv & \int_{0}^{t}dt'\sum_{k}|g_{k}|^{2}\left(\bar{N_{k}}+1\right)\cos\left(\omega_{k0}(t-t')\right),\label{eq:Dr' def}\\
D'_{I}(t) & \equiv & \int_{0}^{t}dt'\sum_{k}|g_{k}|^{2}\left(\bar{N_{k}}+1\right)\sin\left(\omega_{k0}(t-t')\right),\label{eq:Di' def}\\
D_{C}(t) & \equiv & \frac{1}{4}\int_{0}^{t}dt'\overline{a(t)a(t')}.\label{eq:Dc def}
\end{eqnarray}
The details of the calculation leading to Eq.(\ref{eq:2nd order eom example})
can be found in Appendix B.

\subsection{Decay of coherence}

With the equation of motion Eq.(\ref{eq:2nd order eom example}),
one can then study various aspects of the dynamics of a Zeeman splitted
atom in the a stochastic B-field interacting with optical cavity.
One aspect of particular interest to AMO and quantum information physics
is how fast the coherence between atomic energy eigenlevels decays
over time \cite{schlosshauer book}, that is, the decay rate of off-diagonal
element(s) of the atomic density matrix in energy eigenbasis (e.g.
$\rho_{01}(t)$).

\subsubsection*{Total decay rate}

To find the decay rate of the off-diagonal element $\rho_{01}(t)$,
we sandwich both sides of Eq.(\ref{eq:2nd order eom example}) with
$\langle0|\ldots|1\rangle$, with the convention that $|0\rangle$
is spin-up and $|1\rangle$ is spin-down. With $\sigma_{+}|0\rangle=0$,
$\sigma_{+}|1\rangle=2|0\rangle$, $\sigma_{-}|0\rangle=2|1\rangle$,
and $\sigma_{-}|1\rangle=0$, it can be shown from Eqs.(\ref{eq:L20 final},
\ref{eq:L02 final}) that 
\begin{eqnarray}
\langle0|\mathcal{L}_{t\,(2,0)}(\rho)|1\rangle & = & 4i\left(D_{I}(t)+D'_{I}(t)\right)\rho_{01}-4\left(D_{R}(t)+D'_{R}(t)\right)\rho_{01},\\
\langle0|\mathcal{L}_{t\,(0,2)}(\rho)|1\rangle & = & -4D_{C}(t)\rho_{01},
\end{eqnarray}
with which we can proceed to have 
\begin{eqnarray}
\frac{d}{dt}\rho_{01}(t) & = & \langle0|\frac{d}{dt}\rho_{t}|1\rangle=\langle0|\mathcal{L}_{t\,(2,0)}(\rho_{t})|1\rangle+\langle0|\mathcal{L}_{t\,(0,2)}(\rho_{t})|1\rangle,\\
\Longrightarrow\qquad\frac{d}{dt}\rho_{01}(t) & = & i\left(4\left(D_{I}(t)+D'_{I}(t)\right)\right)\rho_{01}(t)-4\left(D_{R}(t)+D'_{R}(t)+D_{C}(t)\right)\rho_{01}(t).\nonumber \\
\label{eq:eom off-diagonal}
\end{eqnarray}

We see that the evolution of $\rho_{01}(t)$ is governed by a simple
linear ordinary differential equation Eq.(\ref{eq:eom off-diagonal}),
that is, the evolution of $\rho_{01}(t)$ is decoupled from the other
density matrix elements. The linear ordinary differential equation
Eq.(\ref{eq:eom off-diagonal}) is first order in time, which means
the rate of change in $\rho_{01}(t)$ is simply given by the right-hand
side of the equation. The first term with a pure imaginary prefactor
results in a phase shift of $\rho_{01}(t)$, while the second term
with a real prefactor results in a decay in the amplitude of $\rho_{01}(t)$.
The total decay rate of the coherence $\rho_{01}(t)$ in the presence
of both the optical cavity and the stochastic B-field is thus 
\begin{equation}
D_{total}(t)=4\left(D_{R}(t)+D'_{R}(t)+D_{C}(t)\right).\label{eq:decay rate total}
\end{equation}

\subsubsection*{Decay rate in stochastic B-field alone}

Suppose the two-level atom is subject to an external stochastic B-field
only, that is, in the absence of the optical cavity. This case of
a classical noise scenario is treated systematically in \cite{james2}.
To find the decay rate of $\rho_{01}(t)$ in this case, we follow
the treatment of a single real Gaussian random process in \cite{james2}.
Quoting the results therein, for a stochastic Hamiltonian of the form
\begin{equation}
H_{S}(t)=a(t)\frac{\sigma_{z}}{2},
\end{equation}
where $a(t)$ is a real Gaussian random process, the equation of motion
is (up to second order) 
\begin{equation}
\frac{d}{dt}\rho_{t}=-i\overline{a(t)}\left[\frac{\sigma_{z}}{2},\,\rho_{t}\right]+\left(\overline{a(t)}\int_{0}^{t}dt'\overline{a(t')}-\int_{0}^{t}dt'\overline{a(t)a(t')}\right)\left[\frac{\sigma_{z}}{2},\,\left[\frac{\sigma_{z}}{2},\,\rho_{t}\right]\right].
\end{equation}
Now that we assume a zero mean for the Gaussian random process in
our example, that is, $\overline{a(t)}=0$, we are left with a simplified
equation of motion 
\begin{eqnarray}
\frac{d}{dt}\rho_{t} & = & -\int_{0}^{t}dt'\overline{a(t)a(t')}\left[\frac{\sigma_{z}}{2},\,\left[\frac{\sigma_{z}}{2},\,\rho_{t}\right]\right]\nonumber \\
 & = & -2D_{C}(t)\left(\rho_{t}-\sigma_{z}\rho_{t}\sigma_{z}\right),\label{eq:eom stochastic}
\end{eqnarray}
where we have used the same definition Eq.(\ref{eq:Dc def}) for $D_{C}(t)\equiv\frac{1}{4}\int_{0}^{t}dt'\overline{a(t)a(t')}$
as in previous parts of this paper.

Sandwiching both sides of Eq.(\ref{eq:eom stochastic}) with $\langle0|\ldots|1\rangle$,
we obtain an ordinary differential equation 
\begin{equation}
\frac{d}{dt}\rho_{01}(t)=-4D_{C}(t)\rho_{01}(t),\label{eq:ode stochastic}
\end{equation}
which shows the decay rate of the coherence $\rho_{01}(t)$ in the
stochastic B-field alone is 
\begin{equation}
D_{B-field}(t)=4D_{C}(t).\label{eq:decay rate stochastic}
\end{equation}

\subsubsection*{Decay rate in optical cavity alone}

Suppose the two-level atom is interacting with an optical cavity only,
that is, in the absence of the stochastic B-field. This case of a
system-bath interaction scenario is treated systematically in \cite{yu}.
The Hamiltonian for this case is 
\begin{equation}
H_{SE}(t)=\sigma_{+}\otimes\left(\sum_{k}g_{k}e^{-i\omega_{k0}t}b_{k}\right)+\sigma_{-}\otimes\left(\sum_{k}g_{k}e^{i\omega_{k0}t}b_{k}^{\dagger}\right),\label{eq:Hse alone}
\end{equation}
and the optical cavity is assumed to be initially thermal, 
\begin{equation}
\rho_{E0}=\prod_{k}\left(\frac{1}{Z_{k}}\sum_{m_{k}=0}^{\infty}e^{-m_{k}\beta\omega_{k}}|m_{k}\rangle\langle m_{k}|\right).\label{eq:thermal cavity alone}
\end{equation}
Using the results in \cite{yu} towards Eqs.(\ref{eq:Hse alone},
\ref{eq:thermal cavity alone}), we can show that the second-order
equation of motion for an atom interacting with an optical cavity
is \footnote{This derivation is similar to one in Appendix B, namely Eqs.(\ref{eq:Hse concise form}-\ref{eq:H eff}).}
\begin{eqnarray}
\frac{d}{dt}\rho_{t} & = & -i\,\left[H_{eff}(t),\,\rho_{t}\right]-D_{R}(t)\left(\sigma_{-}\sigma_{+}\rho_{t}+\rho_{t}\sigma_{-}\sigma_{+}-2\sigma_{+}\rho_{t}\sigma_{-}\right)\nonumber \\
 &  & -D'_{R}(t)\left(\sigma_{+}\sigma_{-}\rho_{t}+\rho_{t}\sigma_{+}\sigma_{-}-2\sigma_{-}\rho_{t}\sigma_{+}\right),\label{eq:eom cavity}
\end{eqnarray}
where $H_{eff}(t)\equiv D_{I}(t)\sigma_{-}\sigma_{+}-D'_{I}(t)\sigma_{+}\sigma_{-}$
and we have used the same definitions Eqs.(\ref{eq:Dr def}, \ref{eq:Di def},
\ref{eq:Dr' def}, \ref{eq:Di' def}) for $D_{R}(t)$, $D_{I}(t)$,
$D'_{R}(t)$, and $D'_{I}(t)$ as in previous parts of this paper.

Sandwiching both sides of Eq.(\ref{eq:eom cavity}) with $\langle0|\ldots|1\rangle$,
we obtain an ordinary differential equation for $\rho_{01}(t)$:
\begin{equation}
\frac{d}{dt}\rho_{01}(t)=i\left(4\left(D_{I}(t)+D'_{I}(t)\right)\right)\rho_{01}(t)-4\left(D_{R}(t)+D'_{R}(t)\right)\rho_{01}(t),\label{eq:ode cavity}
\end{equation}
which shows the decay rate of the coherence $\rho_{01}(t)$ due to
atom-cavity interaction alone is 
\begin{equation}
D_{cavity}(t)=4\left(D_{R}(t)+D'_{R}(t)\right).\label{eq:decay rate cavity}
\end{equation}
Note that the first term of Eq.(\ref{eq:ode cavity}) with a pure
imaginary prefactor only results in a phase shift of $\rho_{01}(t)$
but has no effect on its amplitude.

\subsubsection*{Summary}

Comparing Eq.(\ref{eq:decay rate total}) with Eq.(\ref{eq:decay rate stochastic})
and Eq.(\ref{eq:decay rate cavity}), we see that 
\begin{equation}
D_{total}(t)=D_{cavity}(t)+D_{B-field}(t),
\end{equation}
that is, the total decay rate of the coherence between atomic energy
eigenlevels $\rho_{01}(t)$ in the presence of both an optical cavity
and a stochastic B-field is a simple sum of the decay rate due to
the optical cavity alone and that due to the stochastic B-field alone.
In other words, up to second order (i.e. the leading order of decoherence),
the optical cavity and the stochastic B-field neither enhance nor
undermine each other in terms of decoherence effect. In terms of physical
parameters, the total decay rate is 
\begin{eqnarray}
D_{total}(t) & = & 4\int_{0}^{t}dt'\sum_{k}|g_{k}|^{2}\bar{N_{k}}\cos\left(\omega_{k0}(t-t')\right)\nonumber \\
 &  & +4\int_{0}^{t}dt'\sum_{k}|g_{k}|^{2}\left(\bar{N_{k}}+1\right)\cos\left(\omega_{k0}(t-t')\right)+\int_{0}^{t}dt'\overline{a(t)a(t')}.\nonumber \\
\end{eqnarray}

One may further ask if and how the optical cavity and the stochastic
B-field would interfere at higher order(s) in terms of the effects
on the system's dynamics. Higher-order equation(s) of motion would
be needed for such investigation, an example of which can be found
in Appendix A.

\section{Conclusion}

We use a two-dimensional series expansion method to construct a new
master equation formalism, which can properly describe the effects
of both quantum noise and classical noise on a system's ensemble-averaged
reduced dynamics in a unified and consistent way. Such a unified treatment
is of theoretical importance, because conceptually speaking quantum
noise and classical noise are of different natures. Regardless of
empirical implications, it is important to have a theory framework
that can properly deal with the dynamical effects of both. In particular,
this formalism can be used to determine if there is interference between
quantum and classical noises on the system's dynamics and will be
able to capture and describe such interference if there is any (in
a perturbative manner). Interestingly, we find that second-order interference
between quantum and classical noises vanishes identically. This finding
may justify simple additive treatments of quantum and classical noises,
especially in weak coupling and/or short time regimes where second-order
effects dominate. We study the dynamics of a Zeeman-splitted atom
in a stochastic B-field interacting with an optical cavity and calculate
the decay rate of coherence between the atom's energy levels, which
is (up to second order) a simple sum of the decay rate due to the
stochastic B-field alone and that due to the atom-cavity interaction
alone. Further details of higher-order dynamics under quantum and
classical noises can be worked out systematically using Eqs.(\ref{eq:eom inhomo},
\ref{eq:eps total}, \ref{eq:eps m,n}), and the question of higher-order
interference can be further investigated within our formalism. In
the future, we can apply this formalism to more realistic experimental
setups where the effect of quantum noise and that of classical noise
are of comparable (similar) significance and where higher-order contributions
matter.

\subsection*{Acknowledgement}

The author would like to thank Professor Eric Heller for comments
on this project.

\subsection*{Appendix A: Higher-order dynamics}

\subsubsection*{A.1 Third-order equation of motion}

We will examine higher-order terms in the general master equation
to find out about interference. To work out the third-order terms,
we first set $Q=2$ in Eq.(\ref{eq:eom inhomo}) so that we have 
\begin{equation}
\frac{d}{dt}\rho_{t}=\dot{\mathfrak{\mathcal{E}_{t}}}\left(\rho_{t}\right)-\dot{\mathfrak{\mathcal{E}_{t}}}\left(\mathfrak{\mathcal{E}}_{t}\left(\rho_{t}\right)\right)+\dot{\mathfrak{\mathcal{E}_{t}}}\left(\mathfrak{\mathcal{E}}_{t}^{(2)}\left(\rho_{t}\right)\right)-\dot{\mathfrak{\mathcal{E}_{t}}}\left(\mathfrak{\mathcal{E}}_{t}^{(3)}\left(\rho_{0}\right)\right).
\end{equation}
Because the last term is of fourth order, that is, $-\dot{\mathfrak{\mathcal{E}_{t}}}\left(\mathfrak{\mathcal{E}}_{t}^{(3)}\left(\rho_{0}\right)\right)\sim\sum_{q=0}^{4}\mathcal{O}\left(\lambda^{q}\delta^{4-q}\right)$,
it can be neglected for our study of third order effects, and thus
we are left with 
\begin{equation}
\frac{d}{dt}\rho_{t}=\dot{\mathfrak{\mathcal{E}_{t}}}\left(\rho_{t}\right)-\dot{\mathfrak{\mathcal{E}_{t}}}\left(\mathfrak{\mathcal{E}}_{t}\left(\rho_{t}\right)\right)+\dot{\mathfrak{\mathcal{E}_{t}}}\left(\mathfrak{\mathcal{E}}_{t}\left(\mathfrak{\mathcal{E}}_{t}\left(\rho_{t}\right)\right)\right).\label{eq:eom 3rd order}
\end{equation}

Let's examine the right-hand side of Eq.(\ref{eq:eom 3rd order})
term by term to work out third-order contributions. For the first
term of Eq.(\ref{eq:eom 3rd order}), it can be shown that the third
order contributions are 
\begin{eqnarray}
\dot{\mathfrak{\mathcal{E}_{t}}}\left(\rho_{t}\right) & = & 1st\,order\,term+2nd\,order\,term\nonumber \\
 &  & +\lambda^{3}\dot{\mathcal{E}}_{t\,(3,0)}(\rho_{t})+\lambda^{2}\delta\dot{\mathcal{E}}_{t\,(2,1)}(\rho_{t})+\lambda\delta^{2}\dot{\mathcal{E}}_{t\,(1,2)}(\rho_{t})+\delta^{3}\dot{\mathcal{E}}_{t\,(0,3)}(\rho_{t})\nonumber \\
 &  & +higher\,order\,terms.\label{eq:eom 3rd order 1}
\end{eqnarray}
For the second term of Eq.(\ref{eq:eom 3rd order}), it can be shown
that the third order contributions are 
\begin{eqnarray}
-\dot{\mathfrak{\mathcal{E}_{t}}}\left(\mathfrak{\mathcal{E}}_{t}\left(\rho_{t}\right)\right) & = & 2nd\,order\,term\nonumber \\
 &  & +\lambda^{3}\left(-\dot{\mathcal{E}}_{t\,(1,0)}(\mathcal{E}_{t\,(2,0)}(\rho_{t}))-\dot{\mathcal{E}}_{t\,(2,0)}(\mathcal{E}_{t\,(1,0)}(\rho_{t}))\right)\nonumber \\
 &  & +\lambda^{2}\delta\{-\dot{\mathcal{E}}_{t\,(1,0)}(\mathcal{E}_{t\,(1,1)}(\rho_{t}))-\dot{\mathcal{E}}_{t\,(0,1)}(\mathcal{E}_{t\,(2,0)}(\rho_{t}))\nonumber \\
 &  & -\dot{\mathcal{E}}_{t\,(2,0)}(\mathcal{E}_{t\,(0,1)}(\rho_{t}))-\dot{\mathcal{E}}_{t\,(1,1)}(\mathcal{E}_{t\,(1,0)}(\rho_{t}))\}\nonumber \\
 &  & +\lambda\delta^{2}\{-\dot{\mathcal{E}}_{t\,(0,1)}(\mathcal{E}_{t\,(1,1)}(\rho_{t}))-\dot{\mathcal{E}}_{t\,(1,0)}(\mathcal{E}_{t\,(0,2)}(\rho_{t}))\nonumber \\
 &  & -\dot{\mathcal{E}}_{t\,(0,2)}(\mathcal{E}_{t\,(1,0)}(\rho_{t}))-\dot{\mathcal{E}}_{t\,(1,1)}(\mathcal{E}_{t\,(0,1)}(\rho_{t}))\}\nonumber \\
 &  & +\delta^{3}\left(-\dot{\mathcal{E}}_{t\,(0,1)}(\mathcal{E}_{t\,(0,2)}(\rho_{t}))-\dot{\mathcal{E}}_{t\,(0,1)}(\mathcal{E}_{t\,(0,2)}(\rho_{t}))\right)\nonumber \\
 &  & +higher\,order\,terms.\label{eq:eom 3rd order 2}
\end{eqnarray}
For the third term of Eq.(\ref{eq:eom 3rd order}), it can be shown
that the third order contributions are 
\begin{eqnarray}
\dot{\mathfrak{\mathcal{E}_{t}}}\left(\mathfrak{\mathcal{E}}_{t}\left(\mathfrak{\mathcal{E}}_{t}\left(\rho_{t}\right)\right)\right) & = & \lambda^{3}\dot{\mathcal{E}}_{t\,(1,0)}(\mathcal{E}_{t\,(1,0)}(\mathcal{E}_{t\,(1,0)}(\rho_{t})))\nonumber \\
 &  & +\lambda^{2}\delta\{\dot{\mathcal{E}}_{t\,(1,0)}(\mathcal{E}_{t\,(1,0)}(\mathcal{E}_{t\,(0,1)}(\rho_{t})))\nonumber \\
 &  & +\dot{\mathcal{E}}_{t\,(1,0)}(\mathcal{E}_{t\,(0,1)}(\mathcal{E}_{t\,(1,0)}(\rho_{t})))+\dot{\mathcal{E}}_{t\,(0,1)}(\mathcal{E}_{t\,(1,0)}(\mathcal{E}_{t\,(1,0)}(\rho_{t})))\}\nonumber \\
 &  & +\lambda\delta^{2}\{\dot{\mathcal{E}}_{t\,(0,1)}(\mathcal{E}_{t\,(0,1)}(\mathcal{E}_{t\,(1,0)}(\rho_{t})))\nonumber \\
 &  & +\dot{\mathcal{E}}_{t\,(0,1)}(\mathcal{E}_{t\,(1,0)}(\mathcal{E}_{t\,(0,1)}(\rho_{t})))+\dot{\mathcal{E}}_{t\,(1,0)}(\mathcal{E}_{t\,(0,1)}(\mathcal{E}_{t\,(0,1)}(\rho_{t})))\}\nonumber \\
 &  & +\delta^{3}\dot{\mathcal{E}}_{t\,(0,1)}(\mathcal{E}_{t\,(0,1)}(\mathcal{E}_{t\,(0,1)}(\rho_{t})))\nonumber \\
 &  & +higher\,order\,terms.\label{eq:eom 3rd order 3}
\end{eqnarray}

\subsubsection*{A.2 Third-order interference $\mathcal{L}_{t\,(2,1)}(\rho_{t})$}

Collecting all terms with prefactor $\lambda^{2}\delta$ in Eqs.(\ref{eq:eom 3rd order 1},
\ref{eq:eom 3rd order 2}, \ref{eq:eom 3rd order 3}), we have the
expression for $\mathcal{L}_{t\,(2,1)}(\rho_{t})$ in the Master equation:
\begin{eqnarray}
\mathcal{L}_{t\,(2,1)}(\rho_{t}) & = & \dot{\mathcal{E}}_{t\,(2,1)}(\rho_{t})-\dot{\mathcal{E}}_{t\,(1,0)}(\mathcal{E}_{t\,(1,1)}(\rho_{t}))-\dot{\mathcal{E}}_{t\,(0,1)}(\mathcal{E}_{t\,(2,0)}(\rho_{t}))\nonumber \\
 &  & -\dot{\mathcal{E}}_{t\,(2,0)}(\mathcal{E}_{t\,(0,1)}(\rho_{t}))-\dot{\mathcal{E}}_{t\,(1,1)}(\mathcal{E}_{t\,(1,0)}(\rho_{t}))+\dot{\mathcal{E}}_{t\,(1,0)}(\mathcal{E}_{t\,(1,0)}(\mathcal{E}_{t\,(0,1)}(\rho_{t})))\nonumber \\
 &  & +\dot{\mathcal{E}}_{t\,(1,0)}(\mathcal{E}_{t\,(0,1)}(\mathcal{E}_{t\,(1,0)}(\rho_{t})))+\dot{\mathcal{E}}_{t\,(0,1)}(\mathcal{E}_{t\,(1,0)}(\mathcal{E}_{t\,(1,0)}(\rho_{t}))).\label{eq:L21 step1}
\end{eqnarray}

Recall that in Eq.(\ref{eq:L11=00003D=00003D0}) we have shown that
for an arbitrary operator $\rho$
\begin{equation}
\dot{\mathcal{E}}_{t\,(1,1)}(\rho)-\dot{\mathcal{E}}_{t\,(1,0)}\left(\mathcal{E}_{t\,(0,1)}(\rho)\right)-\dot{\mathcal{E}}_{t\,(0,1)}\left(\mathcal{E}_{t\,(1,0)}(\rho)\right)=0.
\end{equation}
Let this arbitrary operator be $\mathcal{E}_{t\,(1,0)}(\rho_{t})$,
in which case we have three terms in Eq.(\ref{eq:L20 step1}) cancelling
out: 
\begin{equation}
-\dot{\mathcal{E}}_{t\,(1,1)}(\mathcal{E}_{t\,(1,0)}(\rho_{t}))+\dot{\mathcal{E}}_{t\,(1,0)}(\mathcal{E}_{t\,(0,1)}(\mathcal{E}_{t\,(1,0)}(\rho_{t})))+\dot{\mathcal{E}}_{t\,(0,1)}(\mathcal{E}_{t\,(1,0)}(\mathcal{E}_{t\,(1,0)}(\rho_{t})))=0.
\end{equation}
We are thus left with 
\begin{eqnarray}
\mathcal{L}_{t\,(2,1)}(\rho_{t}) & = & \dot{\mathcal{E}}_{t\,(2,1)}(\rho_{t})-\dot{\mathcal{E}}_{t\,(2,0)}(\mathcal{E}_{t\,(0,1)}(\rho_{t}))-\dot{\mathcal{E}}_{t\,(1,0)}(\mathcal{E}_{t\,(1,1)}(\rho_{t}))\nonumber \\
 &  & -\dot{\mathcal{E}}_{t\,(0,1)}(\mathcal{E}_{t\,(2,0)}(\rho_{t}))+\dot{\mathcal{E}}_{t\,(1,0)}(\mathcal{E}_{t\,(1,0)}(\mathcal{E}_{t\,(0,1)}(\rho_{t}))).\label{eq:L21 step2}
\end{eqnarray}

Now we can use definition of $\mathcal{E}_{t\,(M,N)}(\rho_{t})$ in
Eq.(\ref{eq:eps m,n}) to evaluate $\mathcal{L}_{t\,(2,1)}(\rho_{t})$.
To facilitate further calculations, let's introduce 
\begin{equation}
\triangle^{(j)}(t)\equiv U_{1,1}^{(j)}(t,0)-U_{1,0}(t,0)U_{0,1}^{(j)}(t,0),
\end{equation}
noting that we have dropped the superscript $j$ for $U_{1,0}^{(j)}(t,0)$
because it does not depend on the stochastic process. It can be shown
that 
\begin{eqnarray}
\triangle^{(j)}(t) & = & (-i)\int_{0}^{t}dt'\left[H_{S}^{(j)}(t')\otimes\mathbb{I}_{E},\,U_{1,0}(t',0)\right]\nonumber \\
 & = & -\int_{0}^{t}dt'\int_{0}^{t'}dt"\left[H_{S}^{(j)}(t')\otimes\mathbb{I}_{E},\,H_{SE}(t")\right],\\
\Longrightarrow\,\lim_{R\rightarrow\infty}\frac{1}{R}\sum_{j=1}^{R}\triangle^{(j)}(t) & = & -\int_{0}^{t}dt'\int_{0}^{t'}dt"\left[\overline{H_{S}(t')}\otimes\mathbb{I}_{E},\,H_{SE}(t")\right]\nonumber \\
 & \equiv & \overline{\triangle}(t).
\end{eqnarray}
With this short-hand notation, it can be shown that the $\mathcal{L}_{t\,(2,1)}(\rho)$
interference term is \footnote{The details of the calculation are mechanical and lengthy, which we
will not show here.} 
\begin{eqnarray}
 &  & \mathcal{L}_{t\,(2,1)}(\rho)\nonumber \\
 & = & (-i)\:Tr_{E}\left(\left[H_{SE}(t),\,\left(\left[\overline{\triangle}(t),\,\rho\otimes\rho_{E0}\right]-Tr_{E}\left(\left[\overline{\triangle}(t),\,\rho\otimes\rho_{E0}\right]\right)\otimes\rho_{E0}\right)\right]\right)\nonumber \\
 & = & i\int_{0}^{t}dt'\int_{0}^{t'}dt"\:Tr_{E}\{\,[H_{SE}(t),\,\nonumber \\
 &  & \left(\left[\left[\overline{H_{S}(t')}\otimes\mathbb{I}_{E},\,H_{SE}(t")\right],\,\rho\otimes\rho_{E0}\right]-Tr_{E}\left(\left[\left[\overline{H_{S}(t')}\otimes\mathbb{I}_{E},\,H_{SE}(t")\right],\,\rho\otimes\rho_{E0}\right]\right)\otimes\rho_{E0}\right)]\,\}.\nonumber \\
\label{eq:L21 step3}
\end{eqnarray}

In general, the system-bath interaction can be written as $H_{SE}(t)=\sum_{n}S_{n}(t)\otimes E_{n}(t)$,
which can be plugged into Eq.(\ref{eq:L21 step3}) to obtain 
\begin{eqnarray}
\mathcal{L}_{t\,(2,1)}(\rho) & = & i\int_{0}^{t}dt'\int_{0}^{t'}dt"\sum_{m}\sum_{n}\{\mathcal{C}_{mn}(t,t")\left[S_{m}(t),\:\left[\overline{H_{S}(t')},\,S_{n}(t")\right]\rho\right]\nonumber \\
 &  & -\mathcal{C}_{nm}(t",t)\left[S_{m}(t),\:\rho\left[\overline{H_{S}(t')},\,S_{n}(t")\right]\right]\},\label{eq:L21 step4}
\end{eqnarray}
where 
\begin{equation}
\mathcal{C}_{jk}(t,t")\equiv Tr_{E}\left(E_{j}(t)E_{k}(t")\rho_{E0}\right)-Tr_{E}\left(E_{j}(t)\rho_{E0}\right)Tr_{E}\left(E_{k}(t")\rho_{E0}\right).
\end{equation}

With Eq.(\ref{eq:L21 step3}) or Eq.(\ref{eq:L21 step4}), we see
that $\mathcal{L}_{t\,(2,1)}(\rho)$ is not identically vanishing.
Thus interference between quantum and classical noises exists in third
order.

\subsection*{Appendix B: Equation of motion for a Zeeman-splitted atom subject
to stochastic B-field and optical cavity}

To work out the equation of motion for the physical example up to
second order, we make use of Eqs.(\ref{eq:L10}-\ref{eq:L11}) to
work out each term in Eq.(\ref{eq:2nd order eom general}).

\subsubsection*{B.1 $\mathcal{L}_{t\,(1,0)}(\rho)$}

The first-order term due to atom-cavity interaction alone is 

\begin{eqnarray}
\mathcal{L}_{t\,(1,0)}(\rho) & = & \dot{\mathcal{E}}_{t\,(1,0)}(\rho)\nonumber \\
 & = & (-i)Tr_{E}\left(\left[H_{SE}(t),\;\rho\otimes\rho_{E0}\right]\right)\nonumber \\
 & = & (-i)\sum_{k}\{g_{k}e^{-i\left(\omega_{k}-\omega_{0}\right)t}Tr_{E}\left(\left[\sigma_{+}b_{k},\;\rho\otimes\rho_{E0}\right]\right)\nonumber \\
 &  & +g_{k}e^{i\left(\omega_{k}-\omega_{0}\right)t}Tr_{E}\left(\left[\sigma_{-}b_{k}^{\dagger},\;\rho\otimes\rho_{E0}\right]\right)\},
\end{eqnarray}
where in the second equality we have used Eq.(\ref{eq:eps 1,0 dot}).
Now we have 
\begin{eqnarray}
Tr_{E}\left(\left[\sigma_{+}b_{k},\;\rho\otimes\rho_{E0}\right]\right) & = & Tr_{E}\left(\sigma_{+}\rho\otimes b_{k}\rho_{E0}-\rho\sigma_{+}\otimes\rho_{E0}b_{k}\right)\nonumber \\
 & = & \sigma_{+}\rho Tr_{E}\left(b_{k}\rho_{E0}\right)-\rho\sigma_{+}Tr_{E}\left(\rho_{E0}b_{k}\right)\nonumber \\
 & = & 0,
\end{eqnarray}
where the last equality results from the cavity being in the thermal
state, 
\begin{eqnarray}
Tr_{E}\left(b_{k}\rho_{E0}\right)=Tr_{E}\left(\rho_{E0}b_{k}\right) & = & Tr_{Ek}\left(\frac{1}{Z_{k}}\sum_{m_{k}=0}^{\infty}e^{-m_{k}\beta\omega_{k}}|m_{k}\rangle\langle m_{k}|b_{k}\right)\nonumber \\
 & = & \frac{1}{Z_{k}}\sum_{m_{k}=0}^{\infty}e^{-m_{k}\beta\omega_{k}}\langle m_{k}|b_{k}|m_{k}\rangle\nonumber \\
 & = & 0.
\end{eqnarray}
Therefore the atom-cavity interaction does not affect the system's
dynamics in first order, 
\begin{equation}
\mathcal{L}_{t\,(1,0)}(\rho)=\dot{\mathcal{E}}_{t\,(1,0)}(\rho)=0.\label{eq:L10=00003D0}
\end{equation}

\subsubsection*{B.2 $\mathcal{L}_{t\,(0,1)}(\rho)$}

The first-order term due to stochasticity of the external B-field
alone is 
\begin{eqnarray}
\mathcal{L}_{t\,(0,1)}(\rho) & = & \dot{\mathcal{E}}_{t\,(0,1)}(\rho)\nonumber \\
 & = & (-i)\left[\overline{H_{S}(t)},\;\rho\right]\nonumber \\
 & = & (-i)\frac{\overline{a(t)}}{2}\left[\sigma_{z},\;\rho\right],
\end{eqnarray}
where $\overline{a(t)}=0$ for all time. Therefore, the stochastic
part of the external B-field does not affect the system's dynamics
in first order, 
\begin{equation}
\mathcal{L}_{t\,(0,1)}(\rho)=\dot{\mathcal{E}}_{t\,(0,1)}(\rho)=0.\label{eq:L01=00003D0}
\end{equation}

\subsubsection*{B.3 $\mathcal{L}_{t\,(2,0)}(\rho)$}

The second-order term due to atom-cavity interaction alone is 
\begin{eqnarray}
\mathcal{L}_{t\,(2,0)}(\rho) & = & \dot{\mathcal{E}}_{t\,(2,0)}(\rho)-\dot{\mathcal{E}}_{t\,(1,0)}\left(\mathcal{E}_{t\,(1,0)}(\rho)\right)\nonumber \\
 & = & \dot{\mathcal{E}}_{t\,(2,0)}(\rho)\nonumber \\
 & = & \frac{\partial}{\partial t}\sum_{m=0}^{2}\lim_{R\rightarrow\infty}\frac{1}{R}\sum_{j=1}^{R}Tr_{E}\left(U_{2-m,0}^{(j)}(t,0)\rho\otimes\rho_{E0}U_{m,0}^{(j)\dagger}(t,0)\right)\nonumber \\
 & = & \frac{\partial}{\partial t}Tr_{E}\left(U_{2,0}(t,0)\rho\otimes\rho_{E0}+U_{1,0}(t,0)\rho\otimes\rho_{E0}U_{1,0}^{\dagger}(t,0)+\rho\otimes\rho_{E0}U_{2,0}^{\dagger}(t,0)\right)\nonumber \\
 & = & Tr_{E}\{\dot{U}_{2,0}(t,0)\rho\otimes\rho_{E0}+\dot{U}_{1,0}(t,0)\rho\otimes\rho_{E0}U_{1,0}^{\dagger}(t,0)\nonumber \\
 &  & +U_{1,0}(t,0)\rho\otimes\rho_{E0}\dot{U}_{1,0}^{\dagger}(t,0)+\rho\otimes\rho_{E0}\dot{U}_{2,0}^{\dagger}(t,0)\},\label{eq:L20 step1}
\end{eqnarray}
where in the second equality we have made use of the vanishing of
$\dot{\mathcal{E}}_{t\,(1,0)}\left(\cdots\right)$, as is already
shown in Eq.(\ref{eq:L10=00003D0}), and in the fourth equalilty we
drop the stochastic averaging because we are dealing with a deterministic
evolution in this case, as is obvious from Eqs.(\ref{eq:U10},\ref{eq:U20}).
Plugging Eqs.(\ref{eq:U10},\ref{eq:U20}) into Eq.(\ref{eq:L20 step1}),
we can carry out further calculation 
\begin{eqnarray}
\mathcal{L}_{t\,(2,0)}(\rho) & = & \dot{\mathcal{E}}_{t\,(2,0)}(\rho)\nonumber \\
 & = & Tr_{E}\{\,-\int_{0}^{t}dt'H_{SE}(t)H_{SE}(t')\rho\otimes\rho_{E0}-\int_{0}^{t}dt'\rho\otimes\rho_{E0}H_{SE}(t')H_{SE}(t)\nonumber \\
 &  & +(-i)H_{SE}(t)\rho\otimes\rho_{E0}(i)\int_{0}^{t}dt'H_{SE}(t')+(-i)\int_{0}^{t}dt'H_{SE}(t')\rho\otimes\rho_{E0}(i)H_{SE}(t)\,\}\nonumber \\
 & = & -\int_{0}^{t}dt\,Tr_{E}\{\,H_{SE}(t)H_{SE}(t')\rho\otimes\rho_{E0}+\rho\otimes\rho_{E0}H_{SE}(t')H_{SE}(t)\nonumber \\
 &  & -H_{SE}(t)\rho\otimes\rho_{E0}H_{SE}(t')-H_{SE}(t')\rho\otimes\rho_{E0}H_{SE}(t)\,\}.
\end{eqnarray}
We see that $\mathcal{L}_{t\,(2,0)}(\rho)$ agrees with the second-order
term in the case of mere quantum noise (i.e. without classical noise)
as in \cite{yu}, as it should.

Following the treatment of \cite{yu}, we use the following notation
to facilitate further derivation 
\begin{eqnarray}
H_{SE}(t) & = & S_{1}\otimes E_{1}(t)+S_{2}\otimes E_{2}(t),\label{eq:Hse concise form}\\
S_{1} & \equiv & \sigma_{+},\\
S_{2} & \equiv & \sigma_{-},\\
E_{1}(t) & \equiv & \sum_{k}g_{k}e^{-i\omega_{k0}t}b_{k},\\
E_{2}(t) & \equiv & \sum_{k}g_{k}e^{i\omega_{k0}t}b_{k}^{\dagger},\\
\omega_{k0} & \equiv & \omega_{k}-\omega_{0},
\end{eqnarray}
in which case the second-order term can be re-written as 
\begin{eqnarray}
\mathcal{L}_{t\,(2,0)}(\rho) & = & -\sum_{m=1,2}\sum_{n=1,2}\int_{0}^{t}dt'\left(\mathcal{C}_{mn}(t,t')\left[S_{m},\:S_{n}\rho\right]-\mathcal{C}_{nm}(t',t)\left[S_{m},\:\rho S_{n}\right]\right),\nonumber \\
\label{eq:L20 step2}\\
\mathcal{C}_{mn}(t,t') & \equiv & Tr_{E}\left(E_{m}(t)E_{n}(t')\rho_{E0}\right).
\end{eqnarray}

We will evaluate the four terms for $m,n=1,2$ respectively in the
following.

For the term $m=n=1$, the first prefactor in Eq.(\ref{eq:L20 step2})
is 
\begin{eqnarray}
\mathcal{C}_{11}(t,t') & = & Tr_{E}\left(E_{1}(t)E_{1}(t')\rho_{E0}\right)\nonumber \\
 & = & \sum_{k}\sum_{k'}g_{k}g_{k'}e^{-i\omega_{k0}t}e^{-i\omega_{k'0}t'}Tr_{E}\left(b_{k}b_{k'}\rho_{E0}\right).\label{eq:C11}
\end{eqnarray}
It is easy to see that the factors $Tr_{E}\left(b_{k}b_{k'}\rho_{E0}\right)$
vanish for the thermal state $\rho_{E0}$. To work it out in detail,
for $k\neq k'$: 
\begin{eqnarray}
Tr_{E}\left(b_{k}b_{k'}\rho_{E0}\right) & = & Tr_{E}\left(b_{k}b_{k'}\prod_{K}\left(\frac{1}{Z_{K}}\sum_{m_{K}=0}^{\infty}e^{-m_{K}\beta\omega_{K}}|m_{K}\rangle\langle m_{K}|\right)\right)\nonumber \\
 & = & \frac{1}{Z_{k}Z_{k'}}Tr_{E_{k}}\left(b_{k}\sum_{m_{k}=0}^{\infty}e^{-m_{k}\beta\omega_{k}}|m_{k}\rangle\langle m_{k}|\right)\nonumber \\
 &  & Tr_{E_{k'}}\left(b_{k'}\sum_{m_{k'}=0}^{\infty}e^{-m_{k'}\beta\omega_{k'}}|m_{k'}\rangle\langle m_{k'}|\right)\nonumber \\
 & = & \frac{1}{Z_{k}Z_{k'}}\left(\sum_{m_{k}=0}^{\infty}e^{-m_{k}\beta\omega_{k}}\langle m_{k}|b_{k}|m_{k}\rangle\right)\nonumber \\
 &  & \left(\sum_{m_{k'}=0}^{\infty}e^{-m_{k'}\beta\omega_{k'}}\langle m_{k'}|b_{k'}|m_{k'}\rangle\right)\nonumber \\
 & = & 0,\label{eq:kk'}
\end{eqnarray}
because all the factors $\langle m_{k}|b_{k}|m_{k}\rangle\propto\langle m_{k}+1|m_{k}\rangle=0$
vanish; for $k=k'$: 
\begin{eqnarray}
Tr_{E}\left(b_{k}b_{k}\rho_{E0}\right) & = & \frac{1}{Z_{k}}Tr_{E_{k}}\left(b_{k}b_{k}\sum_{m_{k}=0}^{\infty}e^{-m_{k}\beta\omega_{k}}|m_{k}\rangle\langle m_{k}|\right)\nonumber \\
 & = & \frac{1}{Z_{k}}\left(\sum_{m_{k}=0}^{\infty}e^{-m_{k}\beta\omega_{k}}\langle m_{k}|b_{k}b_{k}|m_{k}\rangle\right)\nonumber \\
 & = & 0,\label{eq:kk}
\end{eqnarray}
because all the factors $\langle m_{k}|b_{k}b_{k}|m_{k}\rangle\propto\langle m_{k}+2|m_{k}\rangle=0$
vanish. Plugging Eqs.(\ref{eq:kk'},\ref{eq:kk}) back into Eq.(\ref{eq:C11}),
we find that the first prefactor in Eq.(\ref{eq:L20 step2}) vanishes,
$\mathcal{C}_{11}(t,t')=0.$ By the same token, we can show that the
second prefactor in Eq.(\ref{eq:L20 step2}) also vanishes, $\mathcal{C}_{11}(t',t)=0.$
Therefore, the term for $m=n=1$ vanishes.

Similarly, it can be shown that the term for $m=n=2$ vanishes as
well, essentially because the prefactors $\mathcal{C}_{22}(t,t')=\mathcal{C}_{22}(t',t)=0$
vanish, which in turn is due to the vanishing of the factor $Tr_{E}\left(b_{k}^{\dagger}b_{k'}^{\dagger}\rho_{E0}\right)=0$.

Thus we only have to consider the cross terms for $\left(m=1,\,n=2\right)$
and $\left(m=2,\,n=1\right)$ in Eq.(\ref{eq:L20 step2}), which now
reads 
\begin{eqnarray}
\mathcal{L}_{t\,(2,0)}(\rho) & = & -\int_{0}^{t}dt'\{\mathcal{C}_{12}(t,t')\left[S_{1},\:S_{2}\rho\right]-\mathcal{C}_{21}(t',t)\left[S_{1},\:\rho S_{2}\right]\nonumber \\
 &  & +\mathcal{C}_{21}(t,t')\left[S_{2},\:S_{1}\rho\right]-\mathcal{C}_{12}(t',t)\left[S_{2},\:\rho S_{1}\right]\}\nonumber \\
 & = & -\int_{0}^{t}dt'\{\,\mathcal{C}_{12}(t,t')\left(\sigma_{+}\sigma_{-}\rho-\sigma_{-}\rho\sigma_{+}\right)+\mathcal{C}_{12}(t',t)\left(\rho\sigma_{+}\sigma_{-}-\sigma_{-}\rho\sigma_{+}\right)\nonumber \\
 &  & +\mathcal{C}_{21}(t,t')\left(\sigma_{-}\sigma_{+}\rho-\sigma_{+}\rho\sigma_{-}\right)+\mathcal{C}_{21}(t',t)\left(\rho\sigma_{-}\sigma_{+}-\sigma_{+}\rho\sigma_{-}\right)\,\}.\label{eq:L20 step3}
\end{eqnarray}
where in the second equality we have rearrange the order of the terms.
The prefactors are to be evaluated as follows. The first prefactor
is 
\begin{eqnarray}
\mathcal{C}_{12}(t,t') & = & Tr_{E}\left(E_{1}(t)E_{2}(t')\rho_{E0}\right)\nonumber \\
 & = & \sum_{k}\sum_{k'}g_{k}g_{k'}e^{-i\omega_{k0}t}e^{i\omega_{k'0}t'}Tr_{E}\left(b_{k}b_{k'}^{\dagger}\rho_{E0}\right),
\end{eqnarray}
where for $k\neq k'$:
\begin{eqnarray}
Tr_{E}\left(b_{k}b_{k'}^{\dagger}\rho_{E0}\right) & = & \frac{1}{Z_{k}Z_{k'}}Tr_{E_{k}}\left(b_{k}\sum_{m_{k}=0}^{\infty}e^{-m_{k}\beta\omega_{k}}|m_{k}\rangle\langle m_{k}|\right)\nonumber \\
 &  & Tr_{E_{k'}}\left(b_{k'}^{\dagger}\sum_{m_{k'}=0}^{\infty}e^{-m_{k'}\beta\omega_{k'}}|m_{k'}\rangle\langle m_{k'}|\right)\nonumber \\
 & = & \frac{1}{Z_{k}Z_{k'}}\left(\sum_{m_{k}=0}^{\infty}e^{-m_{k}\beta\omega_{k}}\langle m_{k}|b_{k}|m_{k}\rangle\right)\nonumber \\
 &  & \left(\sum_{m_{k'}=0}^{\infty}e^{-m_{k'}\beta\omega_{k'}}\langle m_{k'}|b_{k'}^{\dagger}|m_{k'}\rangle\right)\nonumber \\
 & = & 0,
\end{eqnarray}
and for $k=k'$: 
\begin{eqnarray}
Tr_{E}\left(b_{k}b_{k}^{\dagger}\rho_{E0}\right) & = & \frac{1}{Z_{k}}Tr_{E_{k}}\left(b_{k}b_{k}^{\dagger}\sum_{m_{k}=0}^{\infty}e^{-m_{k}\beta\omega_{k}}|m_{k}\rangle\langle m_{k}|\right)\nonumber \\
 & = & \frac{1}{Z_{k}}\left(\sum_{m_{k}=0}^{\infty}e^{-m_{k}\beta\omega_{k}}\langle m_{k}|b_{k}b_{k}^{\dagger}|m_{k}\rangle\right)\nonumber \\
 & = & \frac{1}{Z_{k}}\left(\sum_{m_{k}=0}^{\infty}e^{-m_{k}\beta\omega_{k}}\langle m_{k}|\left(b_{k}^{\dagger}b_{k}+\mathbb{I}\right)|m_{k}\rangle\right)\nonumber \\
 & = & \frac{1}{Z_{k}}\left(\sum_{m_{k}=0}^{\infty}e^{-m_{k}\beta\omega_{k}}\langle m_{k}|b_{k}^{\dagger}b_{k}|m_{k}\rangle+\sum_{m_{k}=0}^{\infty}e^{-m_{k}\beta\omega_{k}}\right)\nonumber \\
 & = & \bar{N_{k}}+1,
\end{eqnarray}
with the average/expected occupation number in the $k$-th mode of
the bath being 
\begin{equation}
\bar{N_{k}}\equiv Tr_{E}\left(b_{k}^{\dagger}b_{k}\rho_{E0}\right)=\frac{1}{Z_{k}}\sum_{m_{k}=0}^{\infty}e^{-m_{k}\beta\omega_{k}}\langle m_{k}|b_{k}^{\dagger}b_{k}|m_{k}\rangle;
\end{equation}
therefore we have 
\begin{eqnarray}
\mathcal{C}_{12}(t,t') & = & \sum_{k}\sum_{k'}g_{k}g_{k'}e^{-i\omega_{k0}t}e^{i\omega_{k'0}t'}Tr_{E}\left(b_{k}b_{k'}^{\dagger}\rho_{E0}\right)\nonumber \\
 & = & \sum_{k}|g_{k}|^{2}e^{-i\omega_{k0}(t-t')}Tr_{E}\left(b_{k}b_{k}^{\dagger}\rho_{E0}\right)\nonumber \\
 & = & \sum_{k}|g_{k}|^{2}\left(\bar{N_{k}}+1\right)e^{-i\omega_{k0}(t-t')}\nonumber \\
 & = & \sum_{k}|g_{k}|^{2}\left(\bar{N_{k}}+1\right)\left(\cos\left(\omega_{k0}(t-t')\right)-i\sin\left(\omega_{k0}(t-t')\right)\right).
\end{eqnarray}
Similarly, the second prefactor is 
\begin{eqnarray}
\mathcal{C}_{12}(t',t) & = & Tr_{E}\left(E_{1}(t')E_{2}(t)\rho_{E0}\right)\nonumber \\
 & = & \sum_{k}\sum_{k'}g_{k}g_{k'}e^{-i\omega_{k0}t'}e^{i\omega_{k'0}t}Tr_{E}\left(b_{k}b_{k'}^{\dagger}\rho_{E0}\right)\nonumber \\
 & = & \sum_{k}|g_{k}|^{2}e^{i\omega_{k0}(t-t')}Tr_{E}\left(b_{k}b_{k}^{\dagger}\rho_{E0}\right)\nonumber \\
 & = & \sum_{k}|g_{k}|^{2}\left(\bar{N_{k}}+1\right)e^{i\omega_{k0}(t-t')}\nonumber \\
 & = & \sum_{k}|g_{k}|^{2}\left(\bar{N_{k}}+1\right)\left(\cos\left(\omega_{k0}(t-t')\right)+i\sin\left(\omega_{k0}(t-t')\right)\right);
\end{eqnarray}
the third prefactor is 
\begin{eqnarray}
\mathcal{C}_{21}(t,t') & = & Tr_{E}\left(E_{2}(t)E_{1}(t')\rho_{E0}\right)\nonumber \\
 & = & \sum_{k}\sum_{k'}g_{k}g_{k'}e^{i\omega_{k0}t}e^{-i\omega_{k'0}t'}Tr_{E}\left(b_{k}^{\dagger}b_{k'}\rho_{E0}\right)\nonumber \\
 & = & \sum_{k}|g_{k}|^{2}e^{i\omega_{k0}(t-t')}Tr_{E}\left(b_{k}^{\dagger}b_{k}\rho_{E0}\right)\nonumber \\
 & = & \sum_{k}|g_{k}|^{2}\bar{N_{k}}\left(\cos\left(\omega_{k0}(t-t')\right)+i\sin\left(\omega_{k0}(t-t')\right)\right);
\end{eqnarray}
and the fourth prefactor is 
\begin{eqnarray}
\mathcal{C}_{21}(t',t) & = & Tr_{E}\left(E_{2}(t')E_{1}(t)\rho_{E0}\right)\nonumber \\
 & = & \sum_{k}\sum_{k'}g_{k}g_{k'}e^{i\omega_{k0}t'}e^{-i\omega_{k'0}t}Tr_{E}\left(b_{k}^{\dagger}b_{k'}\rho_{E0}\right)\nonumber \\
 & = & \sum_{k}|g_{k}|^{2}e^{-i\omega_{k0}(t-t')}Tr_{E}\left(b_{k}^{\dagger}b_{k}\rho_{E0}\right)\nonumber \\
 & = & \sum_{k}|g_{k}|^{2}\bar{N_{k}}\left(\cos\left(\omega_{k0}(t-t')\right)-i\sin\left(\omega_{k0}(t-t')\right)\right).
\end{eqnarray}

For convenience, let's introduce the following definitions: 
\begin{eqnarray}
D_{R}(t) & \equiv & \int_{0}^{t}dt'\sum_{k}|g_{k}|^{2}\bar{N_{k}}\cos\left(\omega_{k0}(t-t')\right),\label{eq:Dr def-1}\\
D_{I}(t) & \equiv & \int_{0}^{t}dt'\sum_{k}|g_{k}|^{2}\bar{N_{k}}\sin\left(\omega_{k0}(t-t')\right),\label{eq:Di def-1}\\
D'_{R}(t) & \equiv & \int_{0}^{t}dt'\sum_{k}|g_{k}|^{2}\left(\bar{N_{k}}+1\right)\cos\left(\omega_{k0}(t-t')\right),\label{eq:Dr' def-1}\\
D'_{I}(t) & \equiv & \int_{0}^{t}dt'\sum_{k}|g_{k}|^{2}\left(\bar{N_{k}}+1\right)\sin\left(\omega_{k0}(t-t')\right),\label{eq:Di' def-1}
\end{eqnarray}
as a result of which Eq.(\ref{eq:L20 step3}) can be re-expressed
as 
\begin{eqnarray}
\mathcal{L}_{t\,(2,0)}(\rho) & = & -\{\,\left(D'_{R}(t)-iD'_{I}(t)\right)\left(\sigma_{+}\sigma_{-}\rho-\sigma_{-}\rho\sigma_{+}\right)\nonumber \\
 &  & +\left(D'_{R}(t)+iD'_{I}(t)\right)\left(\rho\sigma_{+}\sigma_{-}-\sigma_{-}\rho\sigma_{+}\right)\nonumber \\
 &  & +\left(D_{R}(t)+iD{}_{I}(t)\right)\left(\sigma_{-}\sigma_{+}\rho-\sigma_{+}\rho\sigma_{-}\right)\nonumber \\
 &  & +\left(D_{R}(t)-iD{}_{I}(t)\right)\left(\rho\sigma_{-}\sigma_{+}-\sigma_{+}\rho\sigma_{-}\right)\nonumber \\
 & = & -D_{R}(t)\left(\sigma_{-}\sigma_{+}\rho+\rho\sigma_{-}\sigma_{+}-2\sigma_{+}\rho\sigma_{-}\right)\nonumber \\
 &  & -D'_{R}(t)\left(\sigma_{+}\sigma_{-}\rho+\rho\sigma_{+}\sigma_{-}-2\sigma_{-}\rho\sigma_{+}\right)\nonumber \\
 &  & -i\,\left(D_{I}(t)\left[\sigma_{-}\sigma_{+},\,\rho\right]-D'_{I}(t)\left[\sigma_{+}\sigma_{-},\,\rho\right]\right).
\end{eqnarray}

Alternatively, put in a compact form, we have 
\begin{eqnarray}
\mathcal{L}_{t\,(2,0)}(\rho) & = & -i\,\left[H_{eff}(t),\,\rho\right]-D_{R}(t)\left(\sigma_{-}\sigma_{+}\rho+\rho\sigma_{-}\sigma_{+}-2\sigma_{+}\rho\sigma_{-}\right)\nonumber \\
 &  & -D'_{R}(t)\left(\sigma_{+}\sigma_{-}\rho+\rho\sigma_{+}\sigma_{-}-2\sigma_{-}\rho\sigma_{+}\right),\label{eq:L20 final}
\end{eqnarray}
where the effective Hamiltonian is defined as 
\begin{equation}
H_{eff}(t)\equiv D_{I}(t)\sigma_{-}\sigma_{+}-D'_{I}(t)\sigma_{+}\sigma_{-}.\label{eq:H eff}
\end{equation}

\subsubsection*{B.4 $\mathcal{L}_{t\,(0,2)}(\rho)$}

The second-order term due to stochasticity of the external B-field
alone is 
\begin{eqnarray}
\mathcal{L}_{t\,(0,2)}(\rho) & = & \dot{\mathcal{E}}_{t\,(0,2)}(\rho)-\dot{\mathcal{E}}_{t\,(0,1)}\left(\mathcal{E}_{t\,(0,1)}(\rho)\right)\nonumber \\
 & = & \dot{\mathcal{E}}_{t\,(0,2)}(\rho)\nonumber \\
 & = & \frac{\partial}{\partial t}\sum_{n=0}^{2}\lim_{R\rightarrow\infty}\frac{1}{R}\sum_{j=1}^{R}Tr_{E}\left(U_{0,2-n}^{(j)}(t,0)\rho\otimes\rho_{E0}U_{0,n}^{(j)\dagger}(t,0)\right)\nonumber \\
 & = & \frac{\partial}{\partial t}\lim_{R\rightarrow\infty}\frac{1}{R}\sum_{j=1}^{R}Tr_{E}\{U_{0,2}^{(j)}(t,0)\rho\otimes\rho_{E0}\nonumber \\
 &  & +U_{0,1}^{(j)}(t,0)\rho\otimes\rho_{E0}U_{0,1}^{(j)\dagger}(t,0)+\rho\otimes\rho_{E0}U_{0,2}^{(j)\dagger}(t,0)\}\nonumber \\
 & = & \lim_{R\rightarrow\infty}\frac{1}{R}\sum_{j=1}^{R}Tr_{E}\{\,\dot{U}_{0,2}^{(j)}(t,0)\rho\otimes\rho_{E0}+\rho\otimes\rho_{E0}\dot{U}_{0,2}^{(j)\dagger}(t,0)\nonumber \\
 &  & +\dot{U}_{0,1}^{(j)}(t,0)\rho\otimes\rho_{E0}U_{0,1}^{(j)\dagger}(t,0)+U_{0,1}^{(j)}(t,0)\rho\otimes\rho_{E0}\dot{U}_{0,1}^{(j)\dagger}(t,0)\,\}.\nonumber \\
\label{eq:L02 step1}
\end{eqnarray}
where in the second equality we have made use of the vanishing of
$\dot{\mathcal{E}}_{t\,(0,1)}\left(\cdots\right)$, as is already
shown in Eq.(\ref{eq:L01=00003D0}). 

To facilitate further calculation, let's first evaluate $U_{0,1}^{(j)}(t,0)$
and $U_{0,2}^{(j)}(t,0)$ in Eqs.(\ref{eq:U01}, \ref{eq:U02}) with
the stochastic part of the Hamiltonian $H_{S}^{(j)}(t)=\frac{a^{(j)}(t)}{2}\sigma_{z}$
given in Eq.(\ref{eq:H random}):
\begin{eqnarray}
U_{0,1}^{(j)}(t,0) & = & (-i)\int_{0}^{t}dt'H_{S}^{(j)}(t')\otimes\mathbb{I}_{E}\nonumber \\
 & = & (-i)\int_{0}^{t}dt'\frac{a^{(j)}(t')}{2}\sigma_{z}\otimes\mathbb{I}_{E}\nonumber \\
 & = & (-\frac{i}{2})\left(\int_{0}^{t}dt'a^{(j)}(t')\right)\sigma_{z}\otimes\mathbb{I}_{E},\label{eq:U01 new}
\end{eqnarray}
\begin{eqnarray}
U_{0,2}^{(j)}(t,0) & = & -\int_{0}^{t}dt'\int_{0}^{t'}dt"\left(H_{S}^{(j)}(t')H_{S}^{(j)}(t")\right)\otimes\mathbb{I}_{E}\nonumber \\
 & = & -\int_{0}^{t}dt'\int_{0}^{t'}dt"\left(\frac{a^{(j)}(t')}{2}\sigma_{z}\frac{a^{(j)}(t")}{2}\sigma_{z}\right)\otimes\mathbb{I}_{E}\nonumber \\
 & = & -\frac{1}{4}\left(\int_{0}^{t}dt'\int_{0}^{t'}dt"a^{(j)}(t')a^{(j)}(t")\right)\sigma_{z}^{2}\otimes\mathbb{I}_{E}\nonumber \\
 & = & -\frac{1}{4}\left(\int_{0}^{t}dt'\int_{0}^{t'}dt"a^{(j)}(t')a^{(j)}(t")\right)\mathbb{I}_{S}\otimes\mathbb{I}_{E},\label{eq:U02 new}
\end{eqnarray}
where in the last equality we have made use of the property of the
Pauli operators $\sigma_{z}^{2}=\mathbb{I}_{S}$ for two-level systems.
Plugging Eqs.(\ref{eq:U01 new}, \ref{eq:U02 new}) into Eq.(\ref{eq:L02 step1}),
we have 
\begin{eqnarray}
\mathcal{L}_{t\,(0,2)}(\rho) & = & \dot{\mathcal{E}}_{t\,(0,2)}(\rho)\nonumber \\
 & = & \lim_{R\rightarrow\infty}\frac{1}{R}\sum_{j=1}^{R}Tr_{E}\{\,-\frac{1}{4}\left(\int_{0}^{t}dt'a^{(j)}(t)a^{(j)}(t')\right)\mathbb{I}_{S}\otimes\mathbb{I}_{E}\left(\rho\otimes\rho_{E0}\right)\nonumber \\
 &  & -\frac{1}{4}\left(\int_{0}^{t}dt'a^{(j)}(t)a^{(j)}(t')\right)\left(\rho\otimes\rho_{E0}\right)\mathbb{I}_{S}\otimes\mathbb{I}_{E}\nonumber \\
 &  & +(-\frac{i}{2})\left(a^{(j)}(t)\right)\sigma_{z}\otimes\mathbb{I}_{E}\left(\rho\otimes\rho_{E0}\right)(\frac{i}{2})\left(\int_{0}^{t}dt'a^{(j)}(t')\right)\sigma_{z}\otimes\mathbb{I}_{E}\nonumber \\
 &  & +(-\frac{i}{2})\left(\int_{0}^{t}dt'a^{(j)}(t')\right)\sigma_{z}\otimes\mathbb{I}_{E}\left(\rho\otimes\rho_{E0}\right)(\frac{i}{2})\left(a^{(j)}(t)\right)\sigma_{z}\otimes\mathbb{I}_{E}\,\}\nonumber \\
 & = & \int_{0}^{t}dt'\lim_{R\rightarrow\infty}\frac{1}{R}\sum_{j=1}^{R}Tr_{E}\{\,-\frac{1}{2}\left(a^{(j)}(t)a^{(j)}(t')\right)\left(\rho\otimes\rho_{E0}\right)\nonumber \\
 &  & +\frac{1}{2}\left(a^{(j)}(t)a^{(j)}(t')\right)\left(\sigma_{z}\rho\sigma_{z}\otimes\rho_{E0}\right)\,\}\nonumber \\
 & = & -\frac{1}{2}\int_{0}^{t}dt'\lim_{R\rightarrow\infty}\frac{1}{R}\sum_{j=1}^{R}\left(a^{(j)}(t)a^{(j)}(t')\right)\left(\rho-\sigma_{z}\rho\sigma_{z}\right)Tr_{E}\left(\rho_{E0}\right)\nonumber \\
 & = & -\frac{1}{2}\int_{0}^{t}dt'\overline{a(t)a(t')}\left(\rho-\sigma_{z}\rho\sigma_{z}\right),
\end{eqnarray}
where $\overline{a(t)a(t')}=\lim_{R\rightarrow\infty}\frac{1}{R}\sum_{j=1}^{R}\left(a^{(j)}(t)a^{(j)}(t')\right)$.

Put in a more compact form, we have 
\begin{equation}
\mathcal{L}_{t\,(0,2)}(\rho)=-2D_{C}(t)\left(\rho-\sigma_{z}\rho\sigma_{z}\right),\label{eq:L02 final}
\end{equation}
where the dephasing rate (due to classical noise) is defined as 
\begin{equation}
D_{C}(t)\equiv\frac{1}{4}\int_{0}^{t}dt'\overline{a(t)a(t')}.\label{eq:Dc def-1}
\end{equation}

\subsubsection*{B.5 Second-order equation of motion}

As has been shown in Eq.(\ref{eq:L11=00003D=00003D0}), the second-order
cross term vanishes, that is, $\mathcal{L}_{t\,(1,1)}(\rho)=0$.

Therefore, with the three terms $\mathcal{L}_{t\,(1,0)}(\rho_{t})$,
$\mathcal{L}_{t\,(0,1)}(\rho_{t})$ and $\mathcal{L}_{t\,(1,1)}(\rho_{t})$
vanishing in Eq.(\ref{eq:2nd order eom general}), and the remaining
two terms $\mathcal{L}_{t\,(2,0)}(\rho_{t})$ and $\mathcal{L}_{t\,(0,2)}(\rho_{t})$
given by Eqs.(\ref{eq:L20 final}, \ref{eq:L02 final}), the equation
of motion for this physical example up to second order is 
\begin{eqnarray}
\frac{d}{dt}\rho_{t} & = & -i\,\left[H_{eff}(t),\,\rho_{t}\right]-D_{R}(t)\left(\sigma_{-}\sigma_{+}\rho_{t}+\rho_{t}\sigma_{-}\sigma_{+}-2\sigma_{+}\rho_{t}\sigma_{-}\right)\nonumber \\
 &  & -D'_{R}(t)\left(\sigma_{+}\sigma_{-}\rho_{t}+\rho_{t}\sigma_{+}\sigma_{-}-2\sigma_{-}\rho_{t}\sigma_{+}\right)-2D_{C}(t)\left(\rho_{t}-\sigma_{z}\rho_{t}\sigma_{z}\right),\nonumber \\
\label{eq:2nd order eom example-1}
\end{eqnarray}
where the effective Hamiltonian is $H_{eff}(t)\equiv D_{I}(t)\sigma_{-}\sigma_{+}-D'_{I}(t)\sigma_{+}\sigma_{-}$
and the prefactors $D_{R}(t)$, $D_{I}(t)$, $D'_{R}(t)$, $D'_{I}(t)$,
and $D_{C}(t)$ are as defined in Eqs.(\ref{eq:Dr def-1}, \ref{eq:Di def-1},
\ref{eq:Dr' def-1}, \ref{eq:Di' def-1}, \ref{eq:Dc def-1}) respectively.

\end{document}